%% file: main.tex
\documentclass[sigconf]{acmart}

\usepackage{enumitem}
\usepackage{listings}
\usepackage{multirow}
\usepackage{graphicx}
\usepackage[normalem]{ulem}
\usepackage{makecell}
\useunder{\uline}{\ul}{}

\AtBeginDocument{%
  }

\newcommand{\sysname}{\textsc{HelpCoach}}

\newcommand\numbercircled[1]{\raisebox{.5pt{\textcircled{\raisebox{-.9pt} {#1}}}}}

\setcopyright{acmlicensed}
\copyrightyear{2018}
\acmYear{2018}
\acmDOI{XXXXXXX.XXXXXXX}
\acmConference[Conference acronym 'XX]{Make sure to enter the correct
  conference title from your rights confirmation email}{June 03--05,
  2018}{Woodstock, NY}
\acmISBN{978-1-4503-XXXX-X/2018/06}

\begin{document}

\title{\sysname{}: Scaffolding Targeted AI Help-Seeking During Problem-Solving}

% \correspondingauthor
% \affiliation{%

\author{Hyoungwook Jin}
\email{jinhw@umich.edu}
\orcid{0000-0003-0253-560X}
\affiliation{%
  \institution{University of Michigan}
  \city{Ann Arbor}
  \state{Michigan}
  \country{USA}
}

\author{Weirui Peng}
\email{weiruip@umich.edu}
\orcid{0009-0001-3417-2447}
\affiliation{%
  \institution{University of Michigan}
  \city{Ann Arbor}
  \state{Michigan}
  \country{USA}
}

\author{Jieun Han}
\email{jieun\_han@kaist.ac.kr}
\orcid{0009-0003-7740-517X}
\affiliation{%
  \institution{KAIST}
  \city{Daejeon}
  \country{Republic of Korea}
}

\author{Q. Vera Liao}
\email{veraliao@umich.edu}
\orcid{0000-0003-4543-7196}
\affiliation{%
  \institution{University of Michigan}
  \city{Ann Arbor}
  \state{Michigan}
  \country{USA}
}

\author{Xu Wang}
\email{xwanghci@umich.edu}
\orcid{0000-0001-5551-0815}
\affiliation{%
  \institution{University of Michigan}
  \city{Ann Arbor}
  \state{Michigan}
  \country{USA}
}

\renewcommand{\shortauthors}{Jin et al.}

\begin{abstract}
\input{section/00_abstract}
\end{abstract}

\begin{CCSXML}
<ccs2012>
   <concept>
       <concept_id>10003120.10003123.10011759</concept_id>
       <concept_desc>Human-centered computing~Empirical studies in interaction design</concept_desc>
       <concept_significance>500</concept_significance>
       </concept>
 </ccs2012>
\end{CCSXML}

\ccsdesc[500]{Human-centered computing~Empirical studies in interaction design}

\keywords{Targeted help-seeking, in situ scaffold design, self-regulation on AI}
\begin{teaserfigure}
 \includegraphics[width=\textwidth]{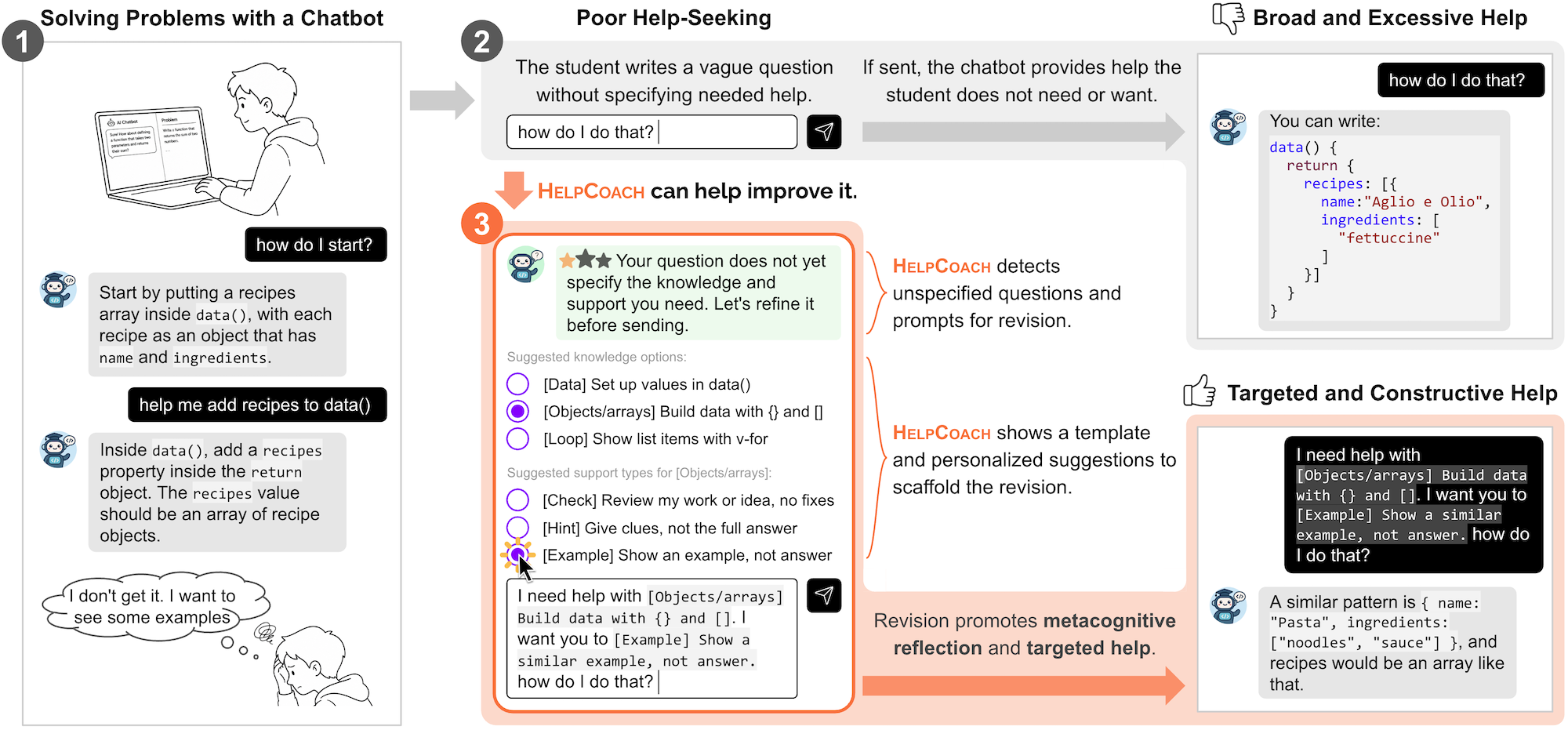}
  \caption{\sysname{} scaffolds targeted help-seeking during problem solving with an AI chatbot. \numbercircled{1}~After receiving general guidance, the student remains confused and wants to see an example. \numbercircled{2}~However, the student asks a vague follow-up question without specifying the relevant knowledge or desired form of scaffold. This unspecified request may elicit a complete solution---and thus more help than the student intended. \numbercircled{3}~In an alternative pathway, if \sysname{} were available, it would detect the unspecified request and prompt the student to revise it before sending. Its template and personalized suggestions help the student articulate the relevant conceptual need and select the desired scaffolding. The revised question encourages metacognitive reflection and elicits targeted, constructive help from the chatbot while leaving the student to adapt the example to the original task.}
  \Description{The figure presents a three-step, left-to-right process showing how HelpCoach helps a student obtain targeted programming assistance from an AI chatbot. Step 1 appears in the left panel, titled "Solving Problems with a Chatbot." A student works on a programming task involving recipe data and asks the chatbot how to begin and how to add recipes to data(). The chatbot explains that the recipes should be represented as an array of objects. A thought bubble shows that the student remains confused and wants to see examples. Step 2 appears across the upper middle and upper right. Under the heading "Poor Help-Seeking," the student types, "how do I do that?" The figure explains that this question does not specify the needed knowledge or form of help. A gray arrow leads to "Broad and Excessive Help," where the chatbot provides a complete code structure for the recipe data. Step 3 occupies the orange-highlighted lower middle and lower right. HelpCoach identifies the question as unspecified and displays suggested knowledge options—data, objects and arrays, and loops—and support options—a check, a hint, or an example. The student selects objects and arrays and requests an example. HelpCoach then generates a revised question that explicitly asks for a similar example rather than an answer. An orange arrow leads to "Targeted and Constructive Help," where the chatbot provides an analogous pasta-recipe object instead of solving the original task. The figure concludes that revising the question promotes metacognitive reflection and targeted help.}
  \label{fig:teaser}
\end{teaserfigure}


\maketitle

\input{section/10_introduction}
\input{section/20_related_work}

\input{section/30_concept}
\input{section/40_study1}
\input{section/50_system}
\input{section/51_techeval}
\input{section/60_study2}
\input{section/70_discussion}
\input{section/80_limitation}
\input{section/90_conclusion}

% \begin{acks}

% \end{acks}

\bibliographystyle{ACM-Reference-Format}
\bibliography{references}

\appendix

%TC:ignore
\section*{Appendix}
\input{appendix/kc_type_classification}
\input{appendix/scaffold_types}
\input{appendix/self_reported_self_regulation_mslq}
%TC:endignore

\end{document}

%% file: section/00_abstract.tex
Students increasingly turn to AI for help with problem-solving, yet too much AI support can undermine learning itself. To benefit from AI, students need to specify the necessary knowledge and scaffold type in their questions. However, they struggle to formulate such targeted questions because they lack metacognitive skills to recognize and select effective help options. We developed \sysname{}, an add-on for chat interfaces that helps students formulate knowledge- and scaffold-specific questions and receive targeted help during problem solving. \sysname{} continuously assesses students' help-seeking performance and prompts students to improve through an adaptive revision template. Whereas prior work has largely taught help-seeking skills apart from learning tasks, \sysname{}'s in situ scaffold enables concrete practice on metacognitive skills and immediate revisions to help-seeking behavior. In a study with 40 college students learning web programming, \sysname{} led to more specific questions during chatbot interactions and greater knowledge retention than pre-task help-seeking training alone.

%% file: section/10_introduction.tex
\section{Introduction}

Help-seeking through generative AI (GenAI) is becoming increasingly common, yet its learning benefits remain inconsistent. As of 2025, Khanmigo had two million users worldwide~\cite{khan2025annual}, and more than 30\% of college students' ChatGPT use involved learning-related tasks~\cite{openai2025college}. Teachers are adopting GenAI chatbots to support personalized help-seeking~\cite{jin2025teachtune, yoo2025teachers}, while students use them as supplementary resources when teacher support is unavailable~\cite{chan2023students, farrokhnia2024swot}. However, prior work reports mixed and often negative learning outcomes when students use GenAI chatbots for problem solving without appropriate interaction design~\cite{stadler2024cognitive, abbas2024harmful, bastani2025generative}. In particular, learning can suffer when students seek direct answers or make little effort to specify the help they need~\cite{fan2025beware, barcaui2025chatgpt, jin2026reliancescope}.

Targeted help-seeking is crucial for effective learning with GenAI chatbots. We characterize \textbf{Targeted help-seeking} as asking questions that specify the \textit{knowledge} needed and the desired form of \textit{scaffold}, so students receive only the help they need to learn, rather than just answers to the problems. Such help-seeking is especially important when interacting with GenAI chatbots trained to provide comprehensive answers to user requests~\cite{macina2023mathdial, bastani2025generative, team2024learnlm}. Targeted help-seeking can support both cognitive and metacognitive learning. By limiting assistance to the necessary scope, students retain room to integrate new knowledge with their existing understanding~\cite{chaiklin2003zone,nelson1981help}. Formulating specific questions also requires students to reflect on their knowledge gaps and learning processes, supporting planning and learner agency~\cite{reeve2011agency, labadze2023role, yan2024promises, yang2025analysing}.

However, students often struggle to seek targeted help because they lack the metacognitive awareness and skills required to ask good questions. Our user studies showed that students ask these targeted questions only 25\% of the time (Section~\ref{sec:study1_quant_finding}). They may not know if they are asking unspecified questions while focusing on problem solving~\cite{tassoti2024assessment, winne2002exploring}. With an unfamiliar knowledge domain, students may find it difficult to define an appropriate help boundary because they cannot readily map the relevant \textit{knowledge components}, the units of knowledge that define a question's scope~\cite{miyake1979ask,pressiey1987cognitive}. They may also be unaware of the range of available \textit{scaffold types}, from concrete examples to high-level hints, or which types best fit their needs~\cite{raz2026knowledge}. This work investigates whether making knowledge components and scaffold types visible during help-seeking in situ can help students recognize, select, and specify these options.

To promote targeted help-seeking, we developed \sysname{}, an extension to chat interfaces that scaffolds students in formulating specific questions during problem solving (Figure~\ref{fig:teaser}, orange box). \sysname{} monitors question specificity in real time, assessing whether each question identifies both a knowledge component and a scaffold type, and provides feedback to keep students aware of targeted help-seeking. When a question is insufficiently specific, \sysname{} prompts students to revise it using a template that provides lists of knowledge components and scaffold types to choose from. It also analyzes students' current code and chat history to recommend relevant knowledge components and literature-informed scaffold types. Through this in situ scaffold, students practice mapping concrete problem-solving needs onto abstract knowledge and choosing effective forms of scaffolding. Its revision scaffold fades when students formulate specific questions in their initial drafts, encouraging spontaneous self-regulation~\cite{van2010scaffolding} while reducing unnecessary interruptions~\cite{kalyuga2009expertise}.

We conducted a between-subjects study with 40 college students to evaluate \sysname{}. Participants received pre-task training on targeted help-seeking and completed three web-programming tasks with a GenAI chatbot in either the \sysname{} condition ($n = 20$) or a \textit{Baseline} condition without in situ scaffold ($n = 20$). Compared with \textit{Baseline}, \sysname{} participants formulated a significantly higher proportion of specific questions ($p=.001$, effect size = 1.262), counting only their initial question drafts before system interventions for that question. This demonstrates that participants learned from \sysname{}'s feedback and internalized self-regulation, rather than merely following system guidance. \sysname{} also led to greater knowledge retention one week later ($p=.005$, Cohen's $d=1.100$), and participants' comments suggested broader benefits for self-regulation and metacognition. Together, these findings suggest that in situ help-seeking scaffolds combining \textit{specificity monitoring}, \textit{templated revision}, and \textit{context-aware recommendations} can improve students' questioning behavior during chatbot interactions and promote deeper knowledge construction to support longer retention, helping address key concerns about ineffective AI use in education.

The novelty of this work lies in introducing \textit{in situ scaffolding for help-seeking}: an interaction approach that helps users regulate \textit{what} assistance to request \textit{while} actively engaged in a task. Our situated scaffolding design creates an opportunity for systems to draw on students' task contexts, enabling context-aware help-option recommendations and fine-grained control over the fading mechanism based on real-time assessment. This combination of context-aware recommendations and performance-based fading in GenAI help-seeking has been underexplored. Our findings provide initial empirical evidence that this approach can significantly improve both question formulation and knowledge gain. We believe our work informs not only how to design learning interactions with AI, but also how users can clarify intent, delegate work, and use GenAI strategically.

Our contributions are summarized as follows:
\begin{itemize}
    \item \sysname{}, a system that scaffolds targeted help-seeking during student-chatbot interactions by providing an adaptive revision template based on real-time monitoring of students' question specificity and code state.

    \item Two technical components of \sysname{}: (1) a question-specificity classifier that identifies whether a student question specifies a knowledge component and scaffold type, and (2) a scoped-exploration module that recommends relevant knowledge components from students' code and chat history.

    \item Empirical evidence from a between-subjects study ($n = 40$) that \sysname{} significantly improves knowledge retention and question specificity during chatbot interactions.

    \item An open dataset collected from our user studies, consisting of student-chatbot interaction logs (5,650 messages), students' problem-solving logs (7,266 code snapshots), and measurements of their knowledge and self-reported self-regulation.
\end{itemize}

%% file: section/20_related_work.tex
\section{Related Work}

We review the theoretical and empirical foundations for pedagogically productive AI use and prior systems that improve learners' help-seeking in interactions with tutors.

\subsection{Characterizing Targeted Help-Seeking}

Help-seeking is a learner-driven and observable behavioral pattern in tutor-assisted problem solving, through which learners solicit knowledge or assistance from a tutor to complete a task~\cite{pintrich2000role, aleven2006toward}. Self-regulation is central to making help-seeking pedagogically meaningful~\cite{newman1994adaptive, zimmerman2002becoming}. Poor self-regulators may seek help over-reliantly, offloading too much cognitive effort to tutors without engaging with the target knowledge~\cite{aleven2000limitations}. In contrast, good self-regulators seek help strategically and autonomously, obtaining adaptive assistance at appropriate moments~\cite{nelson1981help, newman2002self}. Thus, productive help-seeking should be distinguished from help avoidance~\cite{ryan2001avoiding, almeda2017help}: what matters is not simply the quantity or frequency of help-seeking, but the scope and form of the requested help~\cite{nelson1986help, wood1999help, bartholome2006matters}. 

In this work, we characterize targeted help-seeking along two dimensions: knowledge scope and scaffold type~\cite{xiao2025improving}. Learners should identify their knowledge gaps and refine their questions within those gaps, so they can integrate new knowledge into their existing understanding independently and incrementally~\cite{vygotsky1978mind, chaiklin2003zone}. They should also select appropriate scaffold types based on their moment-to-moment cognitive load~\cite{walter2017online, van2005cognitive}, such as requesting worked examples for a directive scaffold or step-by-step explanations for a more facilitative scaffold~\cite{tuovinen1999comparison, zhang2026does}. Articulating these dimensions in questions can support learning in two ways. First, the articulation process can promote metacognition by requiring learners to identify their knowledge gaps and reflect on their learning~\cite{gama2004metacognition}. Second, articulated questions can support cognitive gains by helping learners obtain the intended scope and type of assistance~\cite{sawalha2024analyzing, hou2024effects}. This second function is especially important in learning with GenAI chatbots, which often provide broad or complete assistance beyond what learners explicitly request~\cite{hellas2023exploring, kazemitabaar2024codeaid}.

However, learners often put minimal effort into articulating their questions to GenAI~\cite{grasser1994, aleven2004toward, jin2026reliancescope} and may struggle to identify their knowledge gaps or choose effective scaffold types~\cite{raz2026knowledge}. This difficulty may stem from learners' limited awareness of the knowledge space they are exploring and the available forms of scaffold~\cite{miyake1979ask, pressiey1987cognitive}. Our work investigates whether making knowledge components and scaffold types visible during help-seeking can improve learners' metacognition and help them self-regulate toward more targeted help-seeking.

\subsection{Tutoring Targeted Help-Seeking}

Targeted help-seeking requires metacognitive awareness and skills~\cite{newman2002self, azevedo2005computer_a}. Prior work shows that these competencies can be taught in both human-tutoring and computer-based learning settings~\cite{derry1986designing, schunk1998self, jonassen1993structural}. In in-person tutoring, teachers can promote metacognitive awareness by emphasizing its importance~\cite{schraw1998promoting}, providing checklists for self-monitoring~\cite{king1991effects}, or giving learners autonomy to explore and choose metacognitive strategies~\cite{van2021connecting}. In computer-based learning, researchers have studied adaptive support in intelligent tutoring systems that track learners' knowledge and actions in real time~\cite{anderson1995cognitive}. One approach is open learner modeling~\cite{bull2010open}, which makes system inferences about learners' knowledge states visible through skill meters~\cite{mitrovic2007evaluating, brusilovsky2005engaging, sun2023effects}, predicted misconceptions~\cite{bull2006computer}, or comparisons with expert models~\cite{shahrour2008does}. These transparent self-representations can help learners monitor, reflect on, and control their learning processes~\cite{bull2013open}. Another approach is Help Tutor, which uses production rules to detect unproductive help-seeking behaviors, such as help abuse or quick-answer attempts, and provides direct feedback on proper help use~\cite{roll2007designing, aleven2006toward, roll2011improving}.

Supporting targeted help-seeking with GenAI chatbots poses new challenges. Compared with intelligent tutoring systems, GenAI interactions are more open-ended, diverse, and difficult to classify using simple rules~\cite{azevedo2018using_b, jin2026reliancescope}. GenAI chatbots also occupy mixed perceived roles: they can function as tutors from whom learners seek help, but also as assistants to whom learners delegate cognitive work for efficiency~\cite{azevedo2005computer_a, viberg2026efficiency}. Moreover, students often use GenAI outside the classroom without self-regulatory guardrails, highlighting the need to support transfer from structured training contexts to more open-ended learning settings~\cite{bastani2025generative, darvishi2024impact}.

Prior work has developed LLM prompting techniques for pedagogical support~\cite{xiao2025improving, ma2025should} and instructional materials that train students in help-seeking~\cite{xiao2026transforming}. Building on this work, we explore and extend in situ scaffolding that supports students' self-regulation during cognitive learning activities~\cite{kapoor2026exploring}. Although help-seeking skills are task-agnostic, we examine whether they develop more effectively when students practice them within hands-on, concrete cognitive tasks (e.g., solving programming problems in this research)~\cite{schraw1998promoting, roll2011improving}.

%% file: section/30_concept.tex
\begin{figure*}
    \centering
    \includegraphics[width=\linewidth]{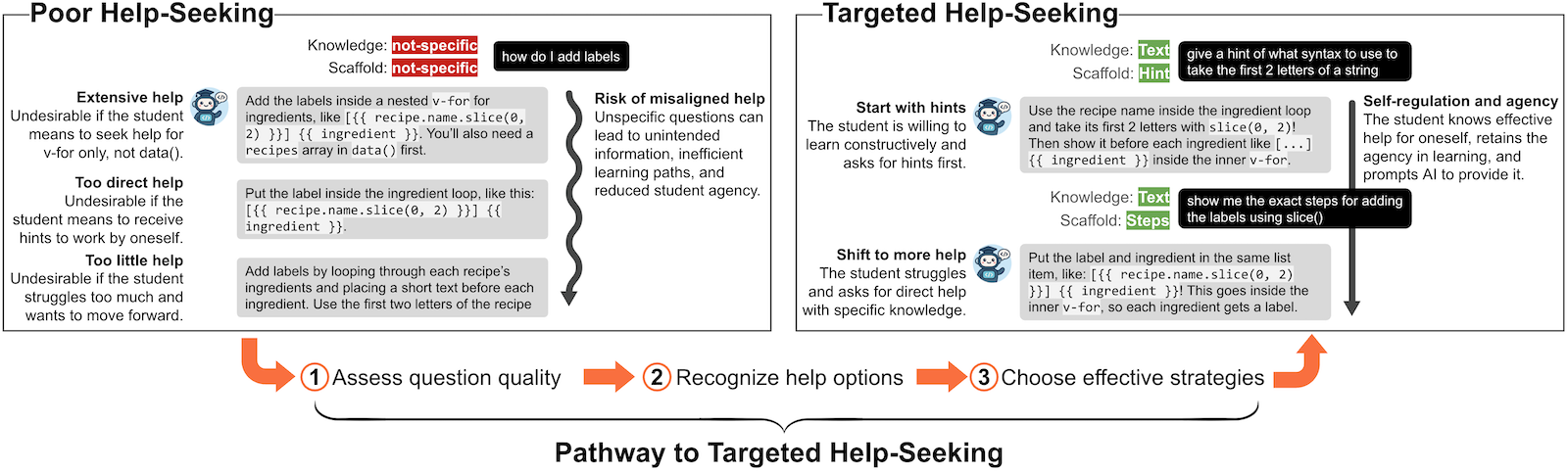}
    \caption{Student-chatbot conversation examples of poor and targeted help-seeking. Poor help-seeking questions can lead chatbots to provide misaligned help, including support that is too extensive, too direct, or insufficient for the student's needs. In contrast, targeted help-seeking allows students to retain agency by specifying both the knowledge they need and the level of support that fits their current learning state. We hypothesize that moving from poor to targeted help-seeking requires three metacognitive skills.}
    \Description{The figure places two bordered conversation panels side by side. The left panel, titled "Poor Help-Seeking," begins with the student question "how do I add labels." Red labels identify both Knowledge and Scaffold as "not-specific." Below are three gray chatbot responses that provide different amounts of assistance. They align with text on the left describing "Extensive help," "Too direct help," and "Too little help." The responses range from adding labels within a nested v-for loop and introducing data slicing, to supplying exact code, to giving only a broad verbal direction. A large downward wavy arrow points toward a warning labeled "Risk of misaligned help." The right panel, titled "Targeted Help-Seeking," shows two successive exchanges connected by a downward arrow. In the first, green labels specify Knowledge as "Text" and Scaffold as "Hint." The student asks for a syntax hint for taking the first two letters, and the chatbot provides a partial example. In the second, Knowledge remains "Text," while Scaffold changes to "Steps." The student requests exact steps, and the chatbot gives more explicit code guidance. Side labels describe this progression as "Start with hints" and "Shift to more help," ending with "Self-regulation and agency." Along the bottom, thick orange arrows connect three numbered skills: "Assess question quality," "Recognize help options," and "Choose effective strategies," with the final arrow curving upward into the targeted-help panel.}
    \label{fig:mechanism}
\end{figure*}

\section{The Pathway to Targeted Help-Seeking}

We aim to improve poor help-seeking behaviors when using GenAI chatbots. Poor help-seeking refers to asking questions that do not specify the relevant scope of knowledge or the desired level of help (see Figure~\ref{fig:mechanism}, left). Students often ask such poor questions by copying and pasting task instructions into chatbot prompts or by posing ambiguous questions that allow chatbots to provide a wide range of random support, including unintended over-help or under-help~\cite{jin2026reliancescope}. Such non-specific help-seeking can hinder learning in several ways. First, students may miss the cognitive effort of translating concrete problem-solving contexts into abstract, reusable knowledge, which can limit transfer to new problems. Second, students may miss opportunities for metacognitive reflection, such as identifying what they do not know and what kind of support they need at a given moment. Third, students may lose agency in their learning because the chatbot determines what help to provide, which may not align with what they want.

Targeted help seeking, in contrast, is a student-driven, self-regulated learning process in which students retain agency over the help they receive by specifying the knowledge and scaffold they want (see Figure~\ref{fig:mechanism}, right). By making the cognitive effort to translate problem-solving needs into specific knowledge, students can reflect on their understanding and steer chatbot responses toward their own learning strategy. This strategy may involve asking for high-level hints that preserve cognitive work when students have sufficient capacity, or direct answers for fact-based knowledge that primarily requires memorization rather than complex reasoning. 

Based on Nelson-LeGall's help-seeking model~\cite{nelson1981help}, targeted help-seeking requires three metacognitive skills (Figure~\ref{fig:mechanism}, bottom): students need to \numbercircled{1}~assess the specificity of their questions, \numbercircled{2}~recognize the types of knowledge and scaffolds they can request, and \numbercircled{3}~decide which forms of help best fit their needs. We call these metacognitive skills the three steps of \textbf{the pathway to targeted help-seeking}. 

Students may initially lack these metacognitive skills and struggle to develop them independently~\cite{azevedo2008externally}. They may also overestimate their self-regulation, believing that they are self-regulating without consistently acting that way~\cite{bjork2013self, avhustiuk2018illusion, winne2002exploring, kruger1999unskilled}. Moreover, because they are still developing domain knowledge, students may struggle to identify the relevant knowledge components and select effective scaffold types for each~\cite{chi1981categorization, aleven2003help}.

In this work, we specifically want students to articulate the knowledge components and scaffold types they need in their questions. Knowledge components are reusable units of knowledge needed to solve target problems, such as knowing that ``the area of a circle is $\pi r^2$''~\cite{koedinger2012knowledge}. Scaffold types refer to different forms of pedagogical support, such as worked examples or direct answers~\cite{van2010scaffolding}. Tables~\ref{tab:knowledge_examples} and \ref{tab:scaffold_examples} show a few examples of knowledge components and scaffold types and how they are articulated in a question.

We focus on knowledge components and scaffold types because, together, they specify the content and form of help. Knowledge components capture \textit{what} the student needs help with and are tied to a specific learning task, whereas scaffold types capture \textit{how} support is provided and are task-agnostic. This combination creates a broad configuration space for specifying help-seeking.

%% file: section/40_study1.tex
\section{Study 1: Classroom Deployment for Data Collection and Analysis}

\begin{figure*}
    \centering
    \includegraphics[width=\linewidth]{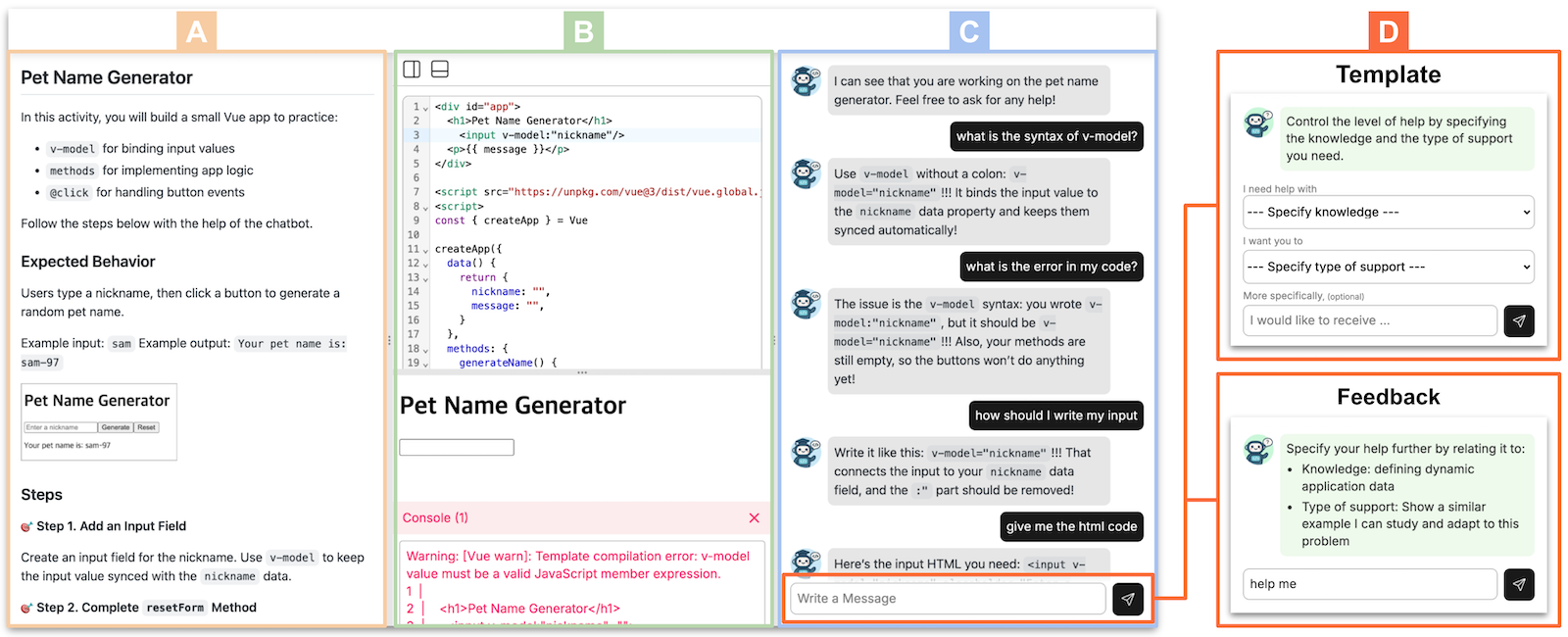}
    \caption{Interface used for the three programming tasks in Study 1. (A) The instruction panel presents the learning objective, target knowledge components, expected outcome, and step-by-step task guidance. (B) Students write code in the editor, and the rendered output appears below; console errors are displayed with the output when present. (C) Students can ask questions to a chatbot, which begins the conversation with the message, ``Feel free to ask for any help!'' (D) In the \textit{Template} and \textit{Feedback} conditions, the message input area includes one of the help-seeking scaffolds.}
    \Description{The figure is arranged as a wide, left-to-right workspace with three adjacent panels outlined in peach (A), green (B), and blue (C), followed by two orange-bordered close-ups (D). Panel A resembles a scrolling worksheet headed "Pet Name Generator," with section headings, bullet points, an example output, and numbered steps. Panel B is vertically divided into a syntax-highlighted code editor, a rendered webpage preview, and a red error console. Panel C shows an alternating chat transcript: chatbot responses appear in light-gray bubbles with a robot icon on the left, whereas student questions appear in black bubbles on the right. The questions progress from asking about syntax and an error to requesting input-writing help and complete HTML code. An orange connector extends from the message box to Panel D. The upper close-up, labeled "Template," contains two drop-down menus and an optional text field. The lower close-up, labeled "Feedback," displays a green box with suggestions above a single text field. Each input area has a black paper-plane send button.}
    \label{fig:interface}
\end{figure*}

As a first step in designing a help-seeking scaffold, we collected students' chatbot interaction logs and examined how specific their questions were. With informed consent and IRB approval, we collected conversation data and problem-solving logs from 44 undergraduate students enrolled in a user interface development course covering Vue.js\footnote{\url{https://vuejs.org/guide/quick-start.html\#using-vue-from-cdn}}. Students used a web-based learning system (Figure~\ref{fig:interface}) to complete three Vue.js programming tasks with a chatbot powered by GPT-5.4-mini.

In addition to collecting interaction logs, we also conducted an initial exploration of two in situ scaffolding designs: \textit{Template} and \textit{Feedback} (Figure~\ref{fig:interface}, D). \textbf{Template} guided students to formulate specific questions by selecting relevant options from dropdown menus. \textbf{Feedback} allowed students to draft a question first, then provided GPT-5.4-mini-generated feedback to support revision before sending the question to the chatbot. Both scaffolds appeared every other message turn, allowing students to seek help independently on alternate turns. Both designs drew on knowledge components identified through cognitive task analysis~\cite{clark2008cognitive} for each programming task and scaffold types adapted from prior scaffolding literature~\cite {van2010scaffolding}. The appendix provides the full set of knowledge components and scaffold types. Students were randomly assigned to one of three conditions: \textit{Template}, \textit{Feedback}, or \textit{NoScaffold}. They also completed surveys about their learning experience and the usability of the two designs. 

The analytic dataset included 1,286 messages from Programming Task 1, 944 from Task 2, and 1,163 from Task 3. We also collected 3,851 code-edit logs and 133 copy-and-paste logs through the learning system. The full dataset, including messages, system logs, LLM prompts, and time spent on each activity, is available in the supplementary materials.

\subsection{Evaluating Question Specificity}

We measured the proportion of specific questions students asked the chatbot. We define \textbf{question specificity} as the proportion of questions within a session that are both knowledge-specific and scaffold-specific (see examples in Table~\ref{tab:specificity_levels}). To capture students' independent help-seeking, we excluded scaffolded message turns: every other turn for the \textit{Template} and \textit{Feedback} conditions.

Two authors independently annotated the specificity of student questions. For each question, they coded knowledge specificity and scaffold specificity separately as \textit{Not specific} or \textit{Specific}. For questions coded as \textit{Specific}, they further identified the relevant knowledge component and/or scaffold type. Examples of annotated questions are shown in Tables~\ref{tab:knowledge_examples} and \ref{tab:scaffold_examples}. The authors annotated 250, 240, and 249 questions from Tasks 1, 2, and 3, respectively, resolving disagreements across four rounds. In the final round, they achieved reliable inter-rater agreement~\cite{graham2012measuring}: 98\% for knowledge components and 89\% for scaffold types.

We then used an LLM-based classifier to annotate the remaining 371, 210, and 311 questions. The classifier operated in two stages: it first determined whether a question was knowledge- or scaffold-specific, then identified the relevant knowledge component or scaffold type for \textit{specific} questions. We selected Gemini-3.5-flash because it performed best among the tested models, achieving accuracies of 0.830 for knowledge specificity and 0.829 for scaffold specificity with the human-annotated dataset. Detailed accuracy results are provided in Section~\ref{sec:question_specificity}.

\input{table/specificity_levels}

\input{table/knowlege_examples}

\input{table/scaffold_examples}

\subsection{Findings}

Students struggled to formulate specific questions during problem solving and often overestimated their performance on asking specific questions. Students' open-ended survey responses highlighted several limitations of \textit{Template} and \textit{Feedback}, including the challenge of balancing metacognitive reflection with cognitive work. We summarize the findings of this exploratory study to inform our system design. We refer to students in the \textit{Template} and \textit{Feedback} conditions as T[1-13] and F[1-16], respectively.

\subsubsection{Students rarely specified both dimensions of help.}
\label{sec:study1_quant_finding}

Consistent with prior work, students often struggled to ask specific questions. Only less than 25\% of questions were specific (Table~\ref{tab:study1_result}). Twenty percent of student questions specified neither a knowledge component nor a scaffold type, leaving the chatbot to determine both the content and form of support. Students also showed limited calibration in assessing their own help-seeking. In the survey, students rated how well they believed their questions specified both knowledge and support type on a 7-point Likert scale. Self-reported specificity was not significantly correlated with observed specificity in any condition (Spearman's correlation: $r_{Baseline}=.080, p_{Baseline}=.778, r_{Template}=-.475, p_{Template}=.101, r_{Feedback}=-.043, p_{Feedback}=.833$); instead, we observed signals for negative correlations even. Together, these results suggest that students may struggle both to assess the specificity of their questions and to translate their help-seeking needs into a specific knowledge component and scaffold type.

\input{table/study1_result}

\subsubsection{The fixed appearance of \textit{Template} and \textit{Feedback} became disruptive when students no longer needed them.}
\label{sec:study1_qual_finding}

The survey responses revealed that students who already viewed themselves as self-regulated learners (T3, T8, F1) reported limited benefit from the scaffolds and instead experienced them as an interruption. Some students felt annoyed by the extra steps required to ask questions later (T3, T8, T10, T12), suggesting that the scaffolds should adapt to students' self-regulation needs over time rather than appear strictly in every other message turn.

Students also remarked that they struggled to fit their questions into \textit{Template}. Students found it limiting when the specific knowledge they wanted to ask about did not fit any dropdown option (T5, T6). Because students had to use the template in every other turn, they also found it difficult to ask follow-up questions fluidly (T9).

Some students commented that \textit{Feedback} was useful in the first few turns but quickly became repetitive (F10). Others thought the effort cost of rewording their questions to make them more specific was too high (F2, F9). Some students also encountered inaccurate feedback that did not match their intended questions (F14, F15) or reappeared even after they incorporated the suggestions (F15).

%% file: table/specificity_levels.tex
% \usepackage{multirow}
\begin{table*}[ht]
\begin{tabular}{llll}
\Xhline{3\arrayrulewidth}
\textbf{Specificity Label} &
  \textbf{Knowledge} &
  \textbf{Scaffold} &
  \textbf{Message Examples} \\ 
\Xhline{3\arrayrulewidth}
Specific &
  O &
  O &
  \begin{tabular}[c]{@{}l@{}}- {[}Loop, Explain{]} "Can you give me an example of how to build an \\array in vue.js" \\ - {[}Objects/arrays, Hint{]} "Do the recipe objects go within data() or \\after it?" \\ - {[}HTML, Check{]} "Have I placed the \textless{}li\textgreater{}\textless{}/li\textgreater and v-for in the \\correct spot" \end{tabular} \\ \hline
\multirow{3}{*}{Not specific} &
  O &
  X &
  \begin{tabular}[c]{@{}l@{}}- {[}Objects/arrays{]} "How do I create an object" \\ - {[}Data{]} "What should be in data" \end{tabular} \\ \cline{2-4}
 &
  X &
  O &
  \begin{tabular}[c]{@{}l@{}}- {[}Steps{]} "Show the right syntax, I don't understand" \\ - {[}Check{]} "What am I doing wrong" \end{tabular} \\ \cline{2-4}
 &
  X &
  X &
  \begin{tabular}[c]{@{}l@{}}- "I don't know how to do that"\\ - "data() \{ return \{ recipes : {[}'Aglio e Olio'{]} \}"\\ - "Error?"\end{tabular} \\ 
\Xhline{3\arrayrulewidth}
\end{tabular}
\caption{Specific questions should clearly state both the needed knowledge component and scaffold type. The text inside the square brackets indicates the specific knowledge component and the scaffold type, if any.}
\label{tab:specificity_levels}
\Description{The table has four columns: specificity label, knowledge, scaffold, and message examples. An “O” indicates that a message specifies the corresponding component, whereas an “X” indicates that it does not. The first row group contains specific questions, which identify both a knowledge component and a scaffold type. The remaining groups contain nonspecific questions that omit either the knowledge component, the scaffold type, or both. Example messages appear in the rightmost column, with square-bracketed labels identifying any knowledge and scaffold categories expressed in each message. Overall, the table illustrates that a question is classified as specific only when it clearly communicates both what the student needs help with and how the student wants to be helped.}
\end{table*}

%% file: table/knowlege_examples.tex
\begin{table*}[ht]
\begin{tabular}{ll}
\Xhline{3\arrayrulewidth}
\textbf{Knowledge Components} &
  \textbf{Message Examples}
   \\
\Xhline{3\arrayrulewidth}
{[}Data{]} Set up values in Vue data() &
  \begin{tabular}[c]{@{}l@{}}- "Should the recipes go inside return or inside data and outside \\return?"\\ - "What can be stored in data()"\end{tabular}
   \\
\hline
{[}Loop{]} Show list items with v-for &
  \begin{tabular}[c]{@{}l@{}}- "How to format a v-for statement and where to put it"\\ - "Have I placed the \textless{}li\textgreater{}\textless{}/li\textgreater and v-for in the correct spot"\end{tabular}
   \\
\hline
{[}Show data{]} Display Vue data with \{\{ \}\} &
  \begin{tabular}[c]{@{}l@{}}- "How to use interpolation to make recipe names headings"\\ - "What does interpolation mean"\end{tabular}
   \\
\hline
{[}Objects/arrays{]} Build data with \{\} and {[}{]} &
  \begin{tabular}[c]{@{}l@{}}- "How should I initialize the array of objects"\\ - "Where would the second list for recipes themselves go? Inside the \\brackets?"\end{tabular}
   \\
\hline
{[}HTML{]} Add bullet points and headings &
  \begin{tabular}[c]{@{}l@{}}- "So then where would I use ul"\\ - "How would I make the bullet points appear"\end{tabular}
   \\
\hline
{[}Text{]} Slice and combine text &
  \begin{tabular}[c]{@{}l@{}}- "How do I get the first 2 characters of a word"\\ - "What do you mean by string slice"\end{tabular}
   \\
\hline
{[}Errors{]} Understand console errors &
  \begin{tabular}[c]{@{}l@{}}- "What is my Uncaught SyntaxError: Unexpected identifier \\'ingredients' in ingredients"\\ - "Explain the console to me"\end{tabular}
   \\
\Xhline{3\arrayrulewidth}
Not specific &
  \begin{tabular}[c]{@{}l@{}}- "Can you give me a sample syntax"\\ - "What is wrong with this"\\ - "How should I go about the label from step 4"\\ - "data() \{ return \{ recipes: {[}'Aglio e Olio'{]} \}"\end{tabular}
   \\ 
\Xhline{3\arrayrulewidth}
\end{tabular}
\caption{Knowledge components for programming task 1 and corresponding example help-seeking messages from Study 1. The messages in the first seven rows are considered \textit{specific} with respect to knowledge components because each articulates a knowledge component beyond copying the task instructions and targets only one knowledge component at a time. The messages in the last row do not satisfy this requirement and are annotated as \textit{not specific}.}
\label{tab:knowledge_examples}
\Description{The table has two columns: knowledge components on the left and example help-seeking messages on the right. Seven specific knowledge categories are listed from top to bottom: setting up values in Vue data(), displaying list items with v-for, displaying Vue data through interpolation, constructing objects and arrays, adding HTML bullets and headings, manipulating text, and understanding console errors. Each category is accompanied by example student messages that target that particular concept. A final row labeled "Not specific" contains broad, ambiguous, or task-repeating messages that do not clearly identify a single underlying knowledge component.}
\end{table*}

%% file: table/scaffold_examples.tex
% \usepackage[normalem]{ulem}
\begin{table*}[ht]
\begin{tabular}{ll}
\Xhline{3\arrayrulewidth}
\textbf{Scaffold Types} &
  \textbf{Message Examples} \\
\Xhline{3\arrayrulewidth}
{[}Steps{]} Tell me the answer or what to do &
  \begin{tabular}[c]{@{}l@{}}- "Show the right syntax, I don't understand"\\ - "Could you give me a step by step"\end{tabular} \\ \hline
{[}Explain{]}: Explain or clarify a concept &
  \begin{tabular}[c]{@{}l@{}}- "What are \textless{}li\textgreater{}and \textless{}ul\textgreater{}"\\ - "Why do I have an error"\end{tabular} \\ \hline
{[}Check{]} Review my work or idea, no fixes &
  \begin{tabular}[c]{@{}l@{}}- "Do I have to add {[}Pa{]} and {[}Ag{]} manually to the ingredients?"\\ - "like this?: \textless{}li v-for recipe in recipes\textgreater{}"\end{tabular} \\ \hline
{[}Hint{]} Give clues, not the full answer &
  \begin{tabular}[c]{@{}l@{}}- "Can you give me a hint on how to make the recipe names \\bold?"\\ - "Should this be a ul or ol"\end{tabular} \\ \hline
{[}Example{]} Show a similar example, not answer &
  \begin{tabular}[c]{@{}l@{}}- "Can you give me an example of the nested for loop"\\ - "Demonstrate the format for v-for"\end{tabular} \\ \hline
{[}Question{]} Ask questions to make me think &
  \begin{tabular}[c]{@{}l@{}}- "Guide me through thought-provoking questions" \\ (Generated example; no Study 1 participant asked this.)\end{tabular} \\ 
\Xhline{3\arrayrulewidth}
Not specific &
  \begin{tabular}[c]{@{}l@{}}- "I am having a parsing issue"\\ - "I think this is wrong"\\ - "How do I do that"\\ - "I don't know how to do that"\end{tabular} \\
\Xhline{3\arrayrulewidth}
\end{tabular}
\caption{Six task-agnostic scaffold types and corresponding example help-seeking messages from Study 1. The messages in the first six rows are considered \textit{specific} with respect to scaffold types because each articulates a scaffold type. The messages in the last row are ambiguous about what help students wanted and are annotated as \textit{not specific}.}
\label{tab:scaffold_examples}
\Description{The table has two columns: scaffold types and corresponding message examples. Six specific scaffold types are arranged vertically: requesting steps or an answer, requesting an explanation, asking for a review without fixes, requesting a hint, requesting a similar example, and asking the chatbot to pose reflective questions. The examples demonstrate increasingly distinct ways of specifying the form of assistance desired. The question-based scaffold includes a generated example because no participant in Study 1 requested this type of support. A final "Not specific" row contains vague messages that do not communicate how the student wants the chatbot to help.}
\end{table*}

%% file: table/study1_result.tex
\begin{table}[ht]
\begin{tabular}{rccc}
\Xhline{3\arrayrulewidth}
\multicolumn{1}{l}{} & \multicolumn{3}{c}{Question Specificity (\%)}                    \\
\multicolumn{1}{l}{} & \textbf{NoScaffold} & \textbf{Template} & \textbf{Feedback} \\
\hline
Task 1               & 26.6               & 26.2             & 18.3             \\
Task 2               & 16.7               & 12.2             & 11.5             \\
Task 3               & 12.8               & 15.4             & 19.1            \\
\Xhline{3\arrayrulewidth}
\end{tabular}
\caption{The proportion of specific questions in each programming task across the conditions. Note that question specificity for \textit{Template} and \textit{Feedback} counts only student questions not guided by either. Average question specificity did not differ statistically significantly across conditions.}
\label{tab:study1_result}
\Description{The table reports question-specificity percentages for three programming tasks. Tasks are arranged in rows, while the NoScaffold, Template, and Feedback conditions appear in columns. Specificity is generally highest in Task 1 for the No Guidance and Template conditions and then decreases in Task 2. In the Feedback condition, specificity is lowest in Task 2 and increases in Task 3. Thus, the relative ordering of conditions varies across tasks rather than showing a consistent advantage for one condition. Average question specificity does not differ significantly across conditions. For the Template and Feedback conditions, the percentages include only questions that students composed without guidance.}
\end{table}

%% file: section/50_system.tex
\section{System: \sysname{} for Scaffolding Targeted Help-Seeking In Situ}

Based on insights and data collection from Study 1, we developed \sysname{} (Figure~\ref{fig:helpcoach}), an add-on to chat interfaces that analyzes the specificity of student questions in real time and scaffolds students to improve them with adaptive support.

Unlike approaches that train targeted help-seeking before or after learning activities~\cite{schworm2012learning, xiao2026transforming, bai2026enhancing}, this scaffold intervenes at the individual message level during cognitive learning activities. We hypothesize that such domain-embedded practice can provide more effective training because self-regulation often begins as a domain-specific metacognitive skill and later develops into a more domain-independent skill~\cite{schraw1998promoting, roll2007designing}.

\subsection{Three Features to Scaffold Targeted Help-Seeking}

\sysname{} supports targeted help-seeking by scaffolding the three metacognitive skills (Figure~\ref{fig:pathway_scaffold}). Throughout the interaction, it monitors question specificity and provides feedback to help students maintain awareness of targeted help-seeking and learn to assess their own questions (Figure~\ref{fig:pathway_scaffold}, \numbercircled{1}). When students struggle to articulate a specific question, \sysname{} provides a template that helps them identify and map their needs to relevant knowledge components and scaffold types (Figure~\ref{fig:pathway_scaffold}, \numbercircled{2}). It also presents a scoped set of options to help students select effective help based on their current understanding (Figure~\ref{fig:pathway_scaffold}, \numbercircled{3}). These scaffolds fade as students demonstrate targeted help-seeking or choose to work independently, supporting the development of spontaneous self-regulation. The three core features of \sysname{} are as follows:

\begin{figure*}
    \centering
    \includegraphics[width=\linewidth]{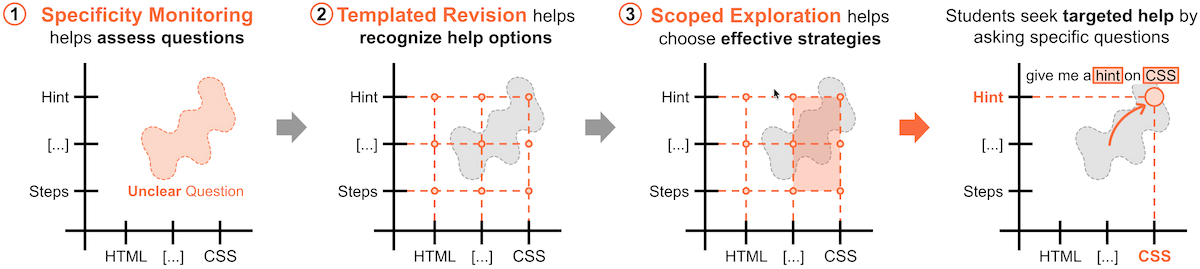}
    \caption{A conceptual overview of how each \sysname{} feature scaffolds the metacognitive skills needed for targeted help-seeking. The grid represents the space of possible help options: the y-axis shows different scaffold types, and the x-axis shows different knowledge components. Unclear questions do not target a specific point in this space; instead, their ambiguity spans a wide range of possible responses, including help that may be unnecessary, ineffective, or unwanted by students. \textit{Template} visualizes this knowledge-scaffold space, while scoped exploration narrows the space students need to consider.}
    \Description{The figure presents four coordinate-style diagrams arranged from left to right and connected by arrows. Each diagram uses the same black axes. The horizontal axis lists knowledge components from HTML through omitted intermediate options to CSS, while the vertical axis lists scaffold types from Steps through intermediate options to Hint. In the first diagram, titled "Specificity Monitoring helps assess questions," a pale-orange irregular shape spans several locations without aligning to a single coordinate. It is labeled "Unclear Question." A gray arrow leads to the second diagram, "Templated Revision helps recognize help options." Here, orange dashed horizontal and vertical lines form a grid, with small circles marking the intersections. A gray irregular region remains visible behind the grid. Another gray arrow leads to "Scoped Exploration helps choose effective strategies." This diagram retains the grid but highlights a smaller orange region on its right side, while the rest of the irregular area remains gray. An orange arrow leads to the final diagram, titled "Students seek targeted help by asking specific questions." Orange dashed guide lines extend from Hint on the vertical axis and CSS on the horizontal axis to a circled point in the upper-right corner. A curved orange arrow points to this coordinate from the remaining gray region. Above it, the example question "give me a hint on CSS" highlights "hint" and "CSS" with orange boxes.}
    \label{fig:pathway_scaffold}
\end{figure*}

\subsubsection{Specificity Monitoring}
\sysname{} continuously assesses the specificity of each student question and displays a brief three-star rating: \textit{specific}, \textit{partially specific}, or \textit{not specific}, where \textit {partially specific} indicates that the question is specific in either knowledge or scaffold alone (Figure~\ref{fig:helpcoach}, \numbercircled{1}). By making question quality visible in each help-seeking attempt, the rating helps students remain aware of targeted help-seeking and learn to evaluate the specificity of their own questions. We use the LLM-based question-specificity classifier developed for Study 1 data analysis to generate these assessments. Unlike the fixed, every-other-turn guidance in Study 1, \sysname{} intervenes with the revision scaffold only when a question is not \textit{specific}. This adaptive timing prevents the unnecessary interruptions that students reported when scaffolds appeared despite their effective help-seeking. It also avoids the expertise reversal effect~\cite{kalyuga2009expertise}, in which support intended to help instead disrupts learners who no longer need it. As students increasingly formulate specific questions, the revision prompt naturally fades, promoting spontaneous self-regulation~\cite{van2010scaffolding}.

\subsubsection{Templated Revision}
Students are asked to revise their questions if they are not specific enough. \sysname{} scaffolds the revision by \textit{Template}, rather than relying solely on facilitative guidance~\cite{shute2008focus} (Figure~\ref{fig:helpcoach}, \numbercircled{2}). The original \textit{Feedback} design lacked structured support for incorporating feedback; in Study 1, students had to interpret textual feedback and revise their questions on their own. As a result, they often skipped feedback without making substantive changes. Because students perceived \textit{Template} as efficient and effective for formulating clear questions, we repurposed it as a revision scaffold. \textit{Template} makes the key dimensions of a specific question visible and helps students recognize available help options rather than recall them~\cite{nielsen1994enhancing}. We expect this guided revision to promote more specific questions in subsequent help-seeking attempts. Students can also opt out by clicking ``Skip guidance this time'' when they prefer to improve questions on their own.

\subsubsection{Scoped Exploration}
\begin{figure*}
    \centering
    \includegraphics[width=\linewidth]{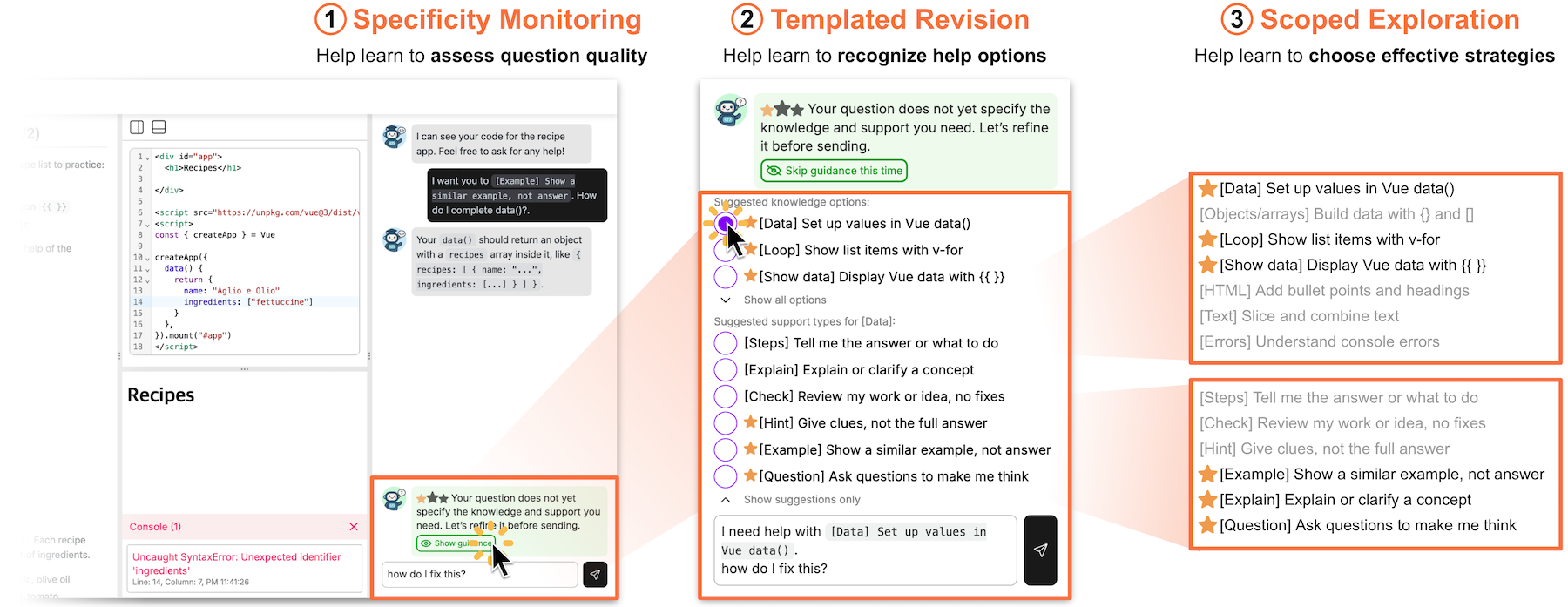}
    \caption{The \sysname{} interface supports targeted help-seeking through three scaffolds. \numbercircled{1}~Specificity monitoring continuously displays a rating of each question's specificity, helping students assess question quality. \numbercircled{2}~Templated revision appears only when a question is insufficiently specific. It shows specific knowledge components and scaffold types so that students can practice mapping their help-seeking needs to them. \numbercircled{3}~Scoped exploration recommends relevant knowledge components and effective scaffold types based on students' chat log and code snapshot. It helps students select appropriate help for their current learning state. Students can expand the suggestions to access the full set of options. The original recommended options are marked with a star.}
    \Description{The figure is organized into three numbered sections with orange headings and black subtitles. The left section shows a programming workspace containing a code editor, a rendered "Recipes" page, a red console error, and a chatbot conversation. At the bottom, an orange-outlined area highlights the question "how do I fix this?" and a pale-green notification displaying one orange and two gray stars. A cursor points to the notification's green "Show guidance" button. The center section enlarges the resulting guidance panel. Beneath the same notification are two lists: suggested knowledge options and suggested support types. Each item has a purple radio button, and several have an orange star. Small controls allow the lists to be expanded or collapsed. The selected knowledge option is inserted into a multiline message field above a black paper-plane send button. Pale-orange bands connect these lists to two enlarged boxes on the right. The upper box shows expanded knowledge options, including Data, Objects, Loop, Show data, HTML, Text, and Errors. The lower box shows support options, including Steps, Check, Hint, Example, Explain, and Question. Starred items appear in darker text with orange stars, whereas additional unstarred options appear in gray.}
    \label{fig:helpcoach}
\end{figure*}

\begin{figure*}
    \centering
    \includegraphics[width=\linewidth]{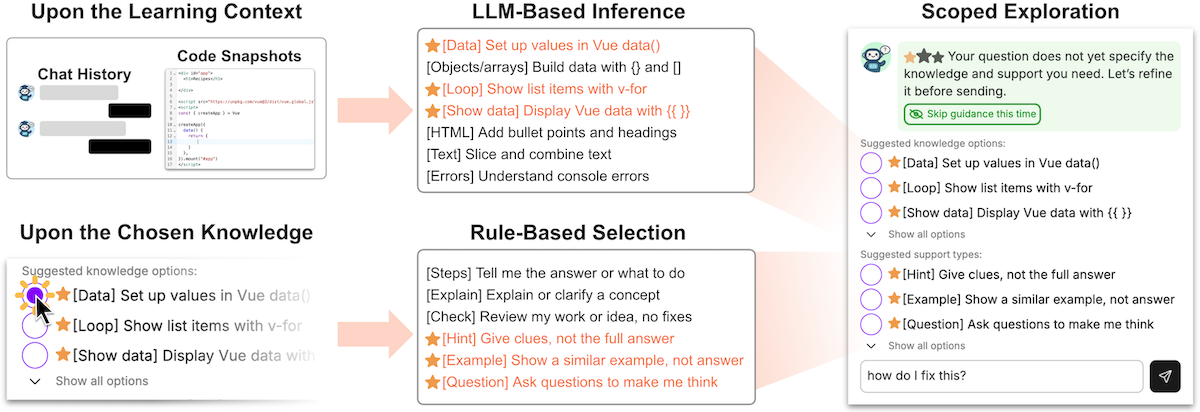}
    \caption{The pipeline for suggesting knowledge components and scaffold types in scoped exploration. We prompted an LLM to infer knowledge components that can help students move forward, considering their chat history and code snapshots. Once students select a knowledge component in the template, \sysname{} shows three scaffold types tailored to it based on the knowledge type (i.e., facts, rules, principles).}
    \Description{The figure presents two parallel left-to-right flows that converge on a single interface. In the upper flow, a box labeled "Upon the Learning Context" contains a schematic chat history with alternating gray and black message bars and a small code-editor screenshot. A peach arrow leads to an "LLM-Based Inference" box listing seven knowledge options. Three options—Data, Loop, and Show data—are highlighted in orange and marked with orange stars; the remaining options appear in black. In the lower flow, "Upon the Chosen Knowledge" shows a cursor selecting Data from a list with purple radio buttons. Another peach arrow leads to a "Rule-Based Selection" box containing six support types. Hint, Example, and Question are highlighted in orange and starred, while Steps, Explain, and Check appear in black. Broad pale-peach arrows connect both result boxes to the "Scoped Exploration" panel on the right. This panel displays a pale-green notification, followed by the three starred knowledge options and the three starred support types, each with a purple radio button. Expand controls appear below both lists. At the bottom is a message field containing "how do I fix this?" and a black paper-plane send button.}
    \label{fig:scoped_exploration}
\end{figure*}

Students receive suggestions for effective knowledge and scaffolds, rather than having to explore all options (Figure~\ref{fig:helpcoach}, \numbercircled{3}). Some students in Study 1 noted that the dropdown menus in \textit{Template} contained too many options and were overwhelming, and that they sometimes struggled to identify which knowledge component to ask about. This issue can become more severe in complex learning tasks, where the space of task-specific knowledge components is larger and more fine-grained. To guide students' search for effective help options, \sysname{} provides \textit{Scoped Exploration} in the template: three knowledge components that are most relevant to students' progress on the current task and three scaffold types suited to the selected component (Figure~\ref{fig:scoped_exploration}). An LLM-based pipeline analyzes students' chat history and code snapshots to identify the most helpful knowledge components to explore. For scaffold-type suggestions, we classify each knowledge component as a fact, rule, or principle and map each category to effective scaffolds based on the KLI framework~\cite{koedinger2012knowledge}. For example, using HTML tags is treated as a rule (i.e., conditional knowledge) because students must select different tags for different purposes; accordingly, \sysname{} suggests \texttt{[Check]}, \texttt{[Hint]}, and \texttt{[Example]} that highlight the relevant conditions when students choose a rule-type knowledge component in the template. If the scoped exploration does not match students' help-seeking needs, they can opt out and choose from the full list. The Appendix (Table~\ref{appendix:kc_type_classification}) provides the classification of knowledge components and their corresponding scaffold-type suggestions.

\subsection{Generalizability of \sysname{}}
While we expect that \sysname{} can generalize across learning domains with well-defined knowledge components, such as mathematics and science, it requires initial inputs that specify the relevant knowledge components and scaffold types. Here, we describe how to determine these inputs.

Knowledge components can be identified through cognitive task analysis, which examines the knowledge and steps used by knowledge experts and novices during task performance~\cite{clark2008cognitive},  learnersourcing, in which students collectively create knowledge components as part of their learning for future students~\cite{jin2024codetree}, or through AI generation~\cite{lee2026knowsim}. The knowledge components should connect task-specific actions to underlying concepts, helping students translate between concrete problems and abstract knowledge. Their granularity depends on the topic and learning objective of the target task. Common student difficulties can also be represented as knowledge components, enabling students to seek targeted help.

Scaffold types can be drawn from prior work on scaffolding~\cite{koedinger2012knowledge}, feedback~\cite{shute2008focus}, or engagement levels~\cite{chi2014icap}. In \sysname{}, we use six scaffolding means from Van de Pol et al.~\cite{van2010scaffolding}: instructing, explaining, feeding back, hints, modeling, and questioning. We order them by cognitive complexity so students can map them onto a continuum and select an appropriate form of support.

%% file: section/51_techeval.tex
\section{Technical Evaluation}

Before evaluating \sysname{} with learners, we first report the performance of its technical components. This technical evaluation serves three purposes: to contextualize how the technical components work, to rule out trivial technical failures that could obscure the findings of the subsequent user study, and to improve the study's reproducibility. Accordingly, our technical evaluation does not include baseline comparisons. We report the performance of three components: the question-specificity classifier, scoped exploration, and chatbot help-response alignment.

\subsection{Question-Specificity Classifier}
\label{sec:question_specificity}

\sysname{} uses an LLM-based classifier to assess each student question. This is the same classifier used to analyze students' question specificity in Study 1. Given the student's question, the chatbot's immediately preceding response, and a snapshot of the student's code when the question was sent, the classifier determines whether the question is specific. It independently outputs knowledge-specificity and scaffold-specificity labels (i.e., \textit{specific}/\textit{not specific}). The supplementary materials include the full system and user prompts.

We report the classifier's percentage accuracy because it performs binary classification. We tested three AI models (Table~\ref{tab:question_specificity_result}) and selected lightweight Flash models because our in situ intervention requires near-instant detection. We compared the model outputs with the human-annotated data from Study 1, including 739 questions. Among the three models, Gemini-3.5-Flash performed the best consistently across the three programming tasks for both knowledge specificity (.830) and scaffold specificity (.829). Given that the average human inter-rater agreement was 98\% for knowledge specificity and 85\% for scaffold specificity, we consider the classifier sufficiently reliable to provide useful real-time feedback to students. We also note that the classifier achieved reasonable accuracy without fine-tuning on a large dataset, indicating its potential scalability to other similar programming tasks.

\input{table/question_specificity_result}

\subsection{Scoped Exploration}

\sysname{}'s scoped exploration identifies three knowledge components that are most relevant to a student's current learning state. We evaluated its performance by comparing its suggestions with human judgments about what knowledge would best support the student. We randomly sampled 150 conversation excerpts from the Study 1 data; each excerpt contained four messages between the student and chatbot, along with the student's code snapshot at the beginning and end of the excerpt. Given each excerpt, human annotators and the scoped exploration identified knowledge components that could help the student move forward or deepen their understanding. Two authors annotated the 150 excerpts through two rounds of conflict resolution, reaching 93\% inter-rater agreement in the second round (n=90).

We tested three LLMs for scoped exploration, as shown in Table~\ref{tab:scoped_suggestion_result}. Performance is reported as the proportion of excerpts in which the three suggested knowledge components included the human-annotated component. GPT-5.6-luna performed best, so we used it in Study 2.

\input{table/scoped_suggestion_result}

\subsection{Help-Response Alignment}
\label{sec:tech_eval_help_reponse_alignment}

\begin{figure}
    \centering
    \includegraphics[width=\linewidth]{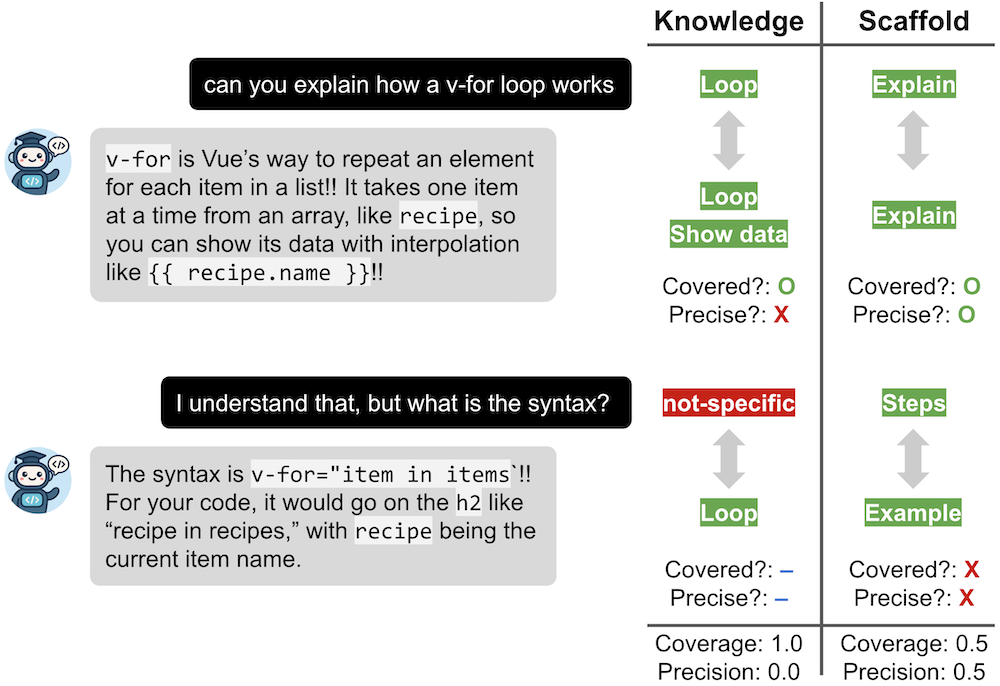}
    \caption{Example calculation of coverage and precision. We calculate coverage and precision separately for knowledge components and scaffold types. In the first message pair, the student asks a specific question requesting an explanation of \texttt{[Loop]}. The chatbot's response \textit{covers} the requested knowledge component but also includes \texttt{[Show data]}, making it not \textit{precise} because it provides more help than requested. For scaffold type, the response is both covered and precise because it provides only the requested form of help (i.e., \texttt{[Explain]}). Coverage and precision are the proportions of covered and precise responses. We calculate coverage and precision only for message pairs in which the student's question is specific, allowing us to assess how well the chatbot aligns its help with specific questions.}
    \Description{The figure is divided into two conversation examples on the left and a two-column evaluation table on the right. Student questions appear in black speech bubbles, and chatbot responses appear in gray boxes beside a robot icon. In the upper example, the student asks, "can you explain how a v-for loop works." The chatbot explains that v-for repeats an element for each item in an array and additionally describes displaying data through interpolation. In the evaluation table, green labels compare the request with the response. Under Knowledge, "Loop" is paired with "Loop" and "Show data," producing a green circle for covered and a red X for precise. Under Scaffold, "Explain" is paired with "Explain," producing green circles for both measures. In the lower example, the student asks for syntax, and the chatbot gives a v-for code example. The requested knowledge is marked "not-specific" in red, so blue dashes replace its coverage and precision results. Under Scaffold, "Steps" is paired with "Example," producing red Xs for both measures. Gray double-headed arrows connect each requested label to the corresponding response label. The bottom row reports Knowledge coverage of 1.0 and precision of 0.0, and Scaffold coverage and precision of 0.5 each.}
    \label{fig:coverage_and_precision}
\end{figure}

We expected students to receive more targeted help when they specified the scope of help in their questions. To examine whether the chatbot provided such targeted help, and thereby incentivized targeted help-seeking, we assessed the alignment between what students requested and what the chatbot provided. We introduce two metrics for quantifying help-response alignment: \textit{coverage} and \textit{precision}. \textbf{Coverage} indicates whether the response includes the information explicitly requested by the student, whereas \textbf{precision} indicates whether the response includes only the requested information (see an example in Figure~\ref{fig:coverage_and_precision}). Precision is a more conservative metric because it assumes coverage. A chatbot with high coverage and precision provides help within the scope the student requested.

We evaluated the coverage and precision of the chatbot used in Study 1. Using human-annotated student questions from our dataset, we asked the chatbot to generate responses, then classified the knowledge components and scaffold types in those responses to calculate the metrics. For this evaluation, we developed another LLM-based multi-label classifier for chatbot responses. Two authors manually annotated 201, 142, and 165 chatbot responses from Study 1 to create a benchmark dataset. Inter-rater reliability, measured as exact-match agreement, was 71\% for knowledge components and 83\% for scaffold types. Because the classification task was multi-label, we report sample-weighted F1 scores. Among the three models tested, Gemini achieved the highest F1 scores of .823 for knowledge components and .768 for scaffold types. The supplementary materials provide a detailed performance report for the chatbot response classifier.

Table~\ref{tab:alignment_result} shows the chatbot's help-response alignment. GPT-5.4-mini powered the chatbot, as in Study 1. Overall, the chatbot showed reasonable coverage and precision for knowledge components, but it often provided more help than students requested, as evidenced by the drop from coverage to precision. In particular, scaffold precision was low: in approximately two-thirds of cases, the chatbot provided more scaffold types than were asked. This over-helping behavior may reduce students' motivation to ask specific questions and weaken \sysname{}'s effect.

To better incentivize targeted help-seeking in Study 2, we revised the chatbot to provide more aligned responses (Table~\ref{tab:alignment_result}). We supplied the chatbot with the classified knowledge component and scaffold type of each student question and programmatically adjusted the system prompt so that it could generate more targeted responses. This approach substantially improved scaffold alignment to above .60, indicating that students would receive targeted help in most cases. The detailed prompt is provided in the supplementary materials.

\input{table/alignment_result}

%% file: table/question_specificity_result.tex
% \usepackage{multirow}
\begin{table*}[ht]
\begin{tabular}{lllllllll}
\Xhline{3\arrayrulewidth}
\multirow{2}{*}{\textbf{Models}} & \multicolumn{4}{l}{\textbf{Knowledge-Specificity}} & \multicolumn{4}{l}{\textbf{Scaffold-Specificity}} \\
 &
  \begin{tabular}[c]{@{}l@{}}Task 1\\ (n=250)\end{tabular} &
  \begin{tabular}[c]{@{}l@{}}Task 2\\ (n=240)\end{tabular} &
  \begin{tabular}[c]{@{}l@{}}Task 3\\ (n=249)\end{tabular} &
  \begin{tabular}[c]{@{}l@{}}Total\\ (n=739)\end{tabular} &
  \begin{tabular}[c]{@{}l@{}}Task 1\\ (n=250)\end{tabular} &
  \begin{tabular}[c]{@{}l@{}}Task 2\\ (n=240)\end{tabular} &
  \begin{tabular}[c]{@{}l@{}}Task 3\\ (n=249)\end{tabular} &
  \begin{tabular}[c]{@{}l@{}}Total\\ (n=739)\end{tabular} \\ 
\Xhline{3\arrayrulewidth}
GPT-5.4-mini                     & .708     & .767     & .759     & .744              & .692     & .771     & .703     & .721             \\
Gemini-3.5-flash                 & .824     & .842     & .823     & \textbf{.830}     & .820     & .850     & .819     & \textbf{.829}    \\
Qwen3-8b                         & .668     & .742     & .759     & .723              & .620     & .679     & .691     & .663             \\
\Xhline{3\arrayrulewidth}
\end{tabular}
\caption{The accuracy of three AI models on specificity classification compared to a human-annotated dataset across three programming tasks. n denotes the number of data points used for the evaluation.}
\label{tab:question_specificity_result}
\Description{The table compares the classification accuracy of GPT-5.4-mini, Gemini-3.5-flash, and Qwen3-8b against human annotations. Models are arranged in rows. Columns are divided into two groups: knowledge-specificity and scaffold-specificity. Each group contains results for Tasks 1 through 3 followed by an overall score, with sample sizes displayed beneath the task headings. Gemini-3.5-flash performs best and remains relatively consistent across tasks for both classification types. GPT-5.4-mini has intermediate overall performance, while Qwen3-8b generally performs lowest, particularly for scaffold-specificity. Accuracy is generally higher for knowledge-specificity than for scaffold-specificity.}
\end{table*}

%% file: table/scoped_suggestion_result.tex
\begin{table*}[ht]
\begin{tabular}{lllll}
\Xhline{3\arrayrulewidth}
\textbf{Models} & \textbf{Task 1 (n=50)} & \textbf{Task 2 (n=50)} & \textbf{Task 3 (n=50)} & \textbf{Total (n=150)} \\
\Xhline{3\arrayrulewidth}
GPT-5.6-luna     & 1.000 & .920 & .920 & \textbf{.947} \\
Gemini-3.5-flash & .980  & .900 & .880 & .920          \\
Qwen3-8b         & .820  & .720 & .740 & .760          \\
\Xhline{3\arrayrulewidth}
\end{tabular}
\caption{Performance of scoped exploration across different LLMs. Performance is measured as the proportion of cases in which the LLM's suggestions included the knowledge component annotated by humans.}
\label{tab:scoped_suggestion_result}
\Description{The table compares scoped-exploration performance across GPT-5.6-luna, Gemini-3.5-flash, and Qwen3-8b. Models are arranged in rows, and columns report performance for Tasks 1, 2, and 3, followed by an overall score. Each task contains 50 cases, for a total of 150 cases. GPT-5.6-luna has the highest overall performance and includes the human-annotated knowledge component in nearly all suggestions. Gemini-3.5-flash performs slightly below GPT-5.6-luna but remains consistently strong. Qwen3-8b performs substantially lower than the other two models. All three models perform best on Task 1 and somewhat less well on Tasks 2 and 3.}
\end{table*}

%% file: table/alignment_result.tex
% \usepackage{multirow}
\begin{table*}[ht]
\begin{tabular}{llllllllll}
\Xhline{3\arrayrulewidth}
                           &         & \multicolumn{4}{l}{\textbf{Knowledge}} & \multicolumn{4}{l}{\textbf{Scaffold}} \\ 
 &
   &
  \begin{tabular}[c]{@{}l@{}}Task 1 \\ (n=72)\end{tabular} &
  \begin{tabular}[c]{@{}l@{}}Task 2 \\ (n=66)\end{tabular} &
  \begin{tabular}[c]{@{}l@{}}Task 3 \\ (n=71)\end{tabular} &
  \begin{tabular}[c]{@{}l@{}}Total  \\ (n=209)\end{tabular} &
  \begin{tabular}[c]{@{}l@{}}Task 1 \\ (n=72)\end{tabular} &
  \begin{tabular}[c]{@{}l@{}}Task 2 \\ (n=66)\end{tabular} &
  \begin{tabular}[c]{@{}l@{}}Task 3 \\ (n=71)\end{tabular} &
  \begin{tabular}[c]{@{}l@{}}Total \\ (n=209)\end{tabular} \\
\Xhline{3\arrayrulewidth}
\multirow{2}{*}{Coverage}  & Study 1 & .972  & .970  & .887  & .943  & .770  & .726  & .704  & .733          \\
                           & Study 2 & .972  & .985  & .959  & \textbf{.971}  & .940  & .821  & .815  & \textbf{.858} \\ \hline
\multirow{2}{*}{Precision} & Study1  & .694  & .773  & .648  & .703           & .330  & .400  & .213  & .310          \\
                           & Study 2 & .792  & .818  & .735  & \textbf{.770}  & .610  & .579  & .585  & \textbf{.607} \\ 
\Xhline{3\arrayrulewidth}
\end{tabular}
\caption{Help-response alignment of the chatbots used in Studies 1 and 2 across the three programming tasks. \textit{n} denotes the number of data points used for evaluation. Coverage and precision indicate the proportion of responses that were covered and precise, respectively.}
\label{tab:alignment_result}
\Description{The table reports chatbot help-response alignment for Studies 1 and 2. Columns are divided into knowledge and scaffold sections, and each section contains results for Tasks 1 through 3 and an overall total. Rows are divided into coverage and precision, with separate rows for each study. The same task-level sample sizes are used in both column sections. Across both studies, knowledge alignment is consistently stronger than scaffold alignment, and coverage is higher than precision. Study 2 generally improves upon Study 1 across tasks and measures. The improvement is especially pronounced for scaffold precision, which is the weakest result in Study 1 but rises substantially in Study 2. Knowledge coverage is high in both studies, while scaffold precision remains the most challenging aspect of help-response alignment.}
\end{table*}

%% file: section/60_study2.tex
\section{Study 2: Lab Study to Evaluate \sysname{}}

We conducted a between-subjects laboratory study with a broader sample of students from our university to evaluate the efficacy of \sysname{}. Participants completed three programming tasks with access to a chatbot for help-seeking. We examined whether \sysname{} supports the pathway to targeted help-seeking and yields the expected gains in knowledge and self-regulation. We compared \sysname{} with a \textit{Baseline} condition that provided pre-task training on help-seeking strategies but no in situ help-seeking scaffold. Specifically, we addressed the following research questions:

\begin{enumerate}[label=\textbf{RQ\arabic*:}]
    \item Does \sysname{} promote more specific questions during chatbot interactions and better retention of this behavior than \textit{Baseline}?
    
    \item Does \sysname{} lead to greater knowledge gains and retention than \textit{Baseline}?
    
    \item Does \sysname{} foster a stronger self-regulatory orientation toward broader learning and greater retention of this orientation than \textit{Baseline}?
    
    \item How does \sysname{} support the three metacognitive skills in the pathway to targeted help-seeking?
    
    \item To what extent does \sysname{}-prompted metacognitive reflection disrupt students' ongoing problem-solving?
\end{enumerate}

\subsection{Procedure}

\begin{figure*}
    \centering
    \includegraphics[width=\linewidth]{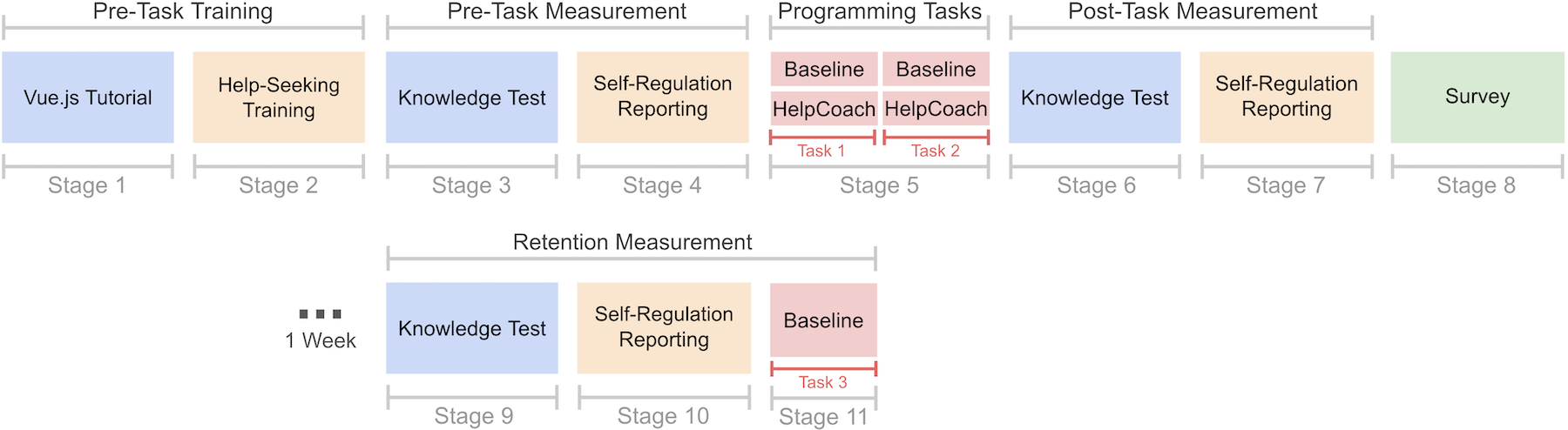}
    \caption{Study 2 procedure. Participants completed pre-task training and measurement, two programming tasks under the \textit{Baseline} or \textit{HelpCoach} condition, and post-task measurement. One week later, they completed retention measures and a third programming task in the \textit{Baseline} condition.}
    \Description{The procedure is presented as a two-row timeline comprising 11 numbered stages. Gray bracket-like lines group the stages into broader phases. The top row begins with pre-task training: a blue Vue.js tutorial box at Stage 1 and a peach help-seeking training box at Stage 2. Pre-task measurement follows, with a blue knowledge test at Stage 3 and peach self-regulation reporting at Stage 4. Stage 5 is a compact pink grid representing two programming tasks. Task 1 and Task 2 form separate columns, each showing Baseline above HelpCoach. Post-task measurement spans Stages 6-8 and includes a blue knowledge test, peach self-regulation reporting, and a green survey. An ellipsis and the label "1 Week" indicate a delay before the lower row. The retention measurement then proceeds through a blue knowledge test at Stage 9, peach self-regulation reporting at Stage 10, and a pink Baseline programming task labeled Task 3 at Stage 11.}
    \label{fig:procedure2}
\end{figure*}

We conducted the study asynchronously and remotely. After enrollment, participants received a link to our web-based learning system and spent a median of 73 minutes completing the main session ($M = 90$ minutes, $SD = 26$ minutes). All participants used the same Vue.js learning interface (Figure~\ref{fig:interface}). The \textit{Baseline} condition used the interface without the in situ help-seeking scaffold, whereas the \textit{HelpCoach} condition had \sysname{}. The study comprised 11 stages (Figure~\ref{fig:procedure2}). Study materials, including questions, chatbot system prompts, and training materials, are available in the supplementary materials.

Participants first completed a Vue.js tutorial (Stage 1 in Figure~\ref{fig:procedure2}) to ensure sufficient foundational knowledge for the programming tasks. To equalize prior exposure to help-seeking strategies, both conditions then received pre-task training on targeted help-seeking (Stage 2). Next, we measured pre-task knowledge and self-regulation (Stages 3 and 4). The knowledge test included six multiple-choice questions and an ``I don't know'' option to discourage random guessing; participants had five minutes to complete it. Participants also completed seven 7-point Likert-scale items measuring metacognitive self-regulation, effort regulation, and help-seeking~\cite{pintrich1991manual}.

Participants then completed two programming tasks (Stage 5), which were identical to the first two tasks in Study 1. Each task required participants to complete different parts of a Vue.js application. For example, the first task focused on completing the data() section, whereas the second focused on completing its HTML template. We used distinct problem settings, skeleton code, and knowledge components across tasks to reduce cross-task learning effects. During the tasks, participants used the interface shown in Figure~\ref{fig:interface}, with a condition-specific help-seeking scaffold in the chat interface. The chatbot was powered by GPT-5.4-mini with a temperature of 1.0, with the prompt explained in Section~\ref{sec:tech_eval_help_reponse_alignment}. Before beginning the tasks, \textit{HelpCoach} participants watched a 1.5-minute tutorial video introducing the interface.

After the programming tasks, participants completed the same knowledge test with a five-minute time limit (Stage 6), followed by the self-regulation measures (Stage 7), and surveys about their learning experience and \sysname{} (Stage 8).

One week later, participants returned for a retention session. They repeated the same knowledge test (Stage 9) and self-regulation measures (Stage 10), then completed the third programming task without an in situ scaffold (Stage 11). This session assessed whether participants retained targeted help-seeking behaviors after \sysname{} was removed.

\subsection{Participants}
We recruited undergraduate and graduate students from our university through departmental advertisements. To ensure that participants could focus on learning Vue.js, we required prior familiarity with basic web programming, including reading and writing HTML, CSS, and JavaScript. We initially recruited 56 participants and excluded those who had participated in Study 1, earned a perfect score on the prior-knowledge test, or sent fewer than three messages to the chatbot during the study. The final analytic sample included 40 participants. Participants provided informed consent and received 30 USD for approximately 1.5 hours of participation. Our institution's IRB approved the study.

In the application form, participants reported their programming and LLM expertise~\cite{feigenspan2012measuring, ma2026not}. On a scale from 1 (very inexperienced) to 7 (very experienced), participants reported an average expertise of $M = 4.0$ ($SD = 1.3$) in web programming and $M = 4.5$ ($SD = 1.4$) in GenAI/LLMs. They reported using GenAI/LLMs for web programming at an average frequency of $M = 4.2$ ($SD = 1.9$), rated from 1 (never) to 7 (daily), and confidence in controlling GenAI/LLMs of $M = 5.3$ ($SD = 1.3$), rated from 1 (very unconfident) to 7 (very confident). Participants were randomly assigned to the \textit{Baseline} or \textit{HelpCoach} condition. 

\subsection{Data Collection}
We collected 1,080 student messages from Programming Task 1, 693 from Task 2, and 646 from Task 3. Twenty participants were assigned to each of the \textit{Baseline} and \textit{HelpCoach} conditions. We also collected 2,704 code-edit records and 111 system interaction logs through the learning system. The full dataset, including student messages, code-edit logs, pre- and post-task knowledge test scores, self-reported self-regulation measures, and time spent at each stage, is available in the supplementary materials.

\subsection{Measures}
\label{sec:study2_measures}

We measured the following outcomes to address our research questions. First, we checked if \sysname{} improves question specificity as intended.
We then tested whether \sysname{} improved knowledge gains and general self-regulation. To explain the results, we also qualitatively examined its pathway to targeted help-seeking using participants' self-reports. Finally, we assessed retention one week after the main session to determine whether \sysname{} supported spontaneous targeted help-seeking after the scaffold was removed, our ultimate goal for students.

\subsubsection{Question specificity (\%)}
This behavioral measure addresses RQ1 by analyzing how students operationalized targeted help-seeking during problem solving. As in Study 1, we used the question-specificity classifier to calculate the proportion of specific questions during the programming tasks in Stages 5 and 11. For the \textit{HelpCoach} condition, we analyzed participants' initial question drafts, before any system intervention or revision, to capture their spontaneous self-regulation.

\subsubsection{Knowledge gain and retention (out of 6)}
This measure directly addresses RQ2. Participants received 1 point for each correct answer to six multiple-choice questions. No points were deducted for incorrect answers or for selecting ``I don't know.'' Participants completed the same test three times, in Stages 3, 6, and 9. We calculated score differences across stages to measure knowledge gains and retention.

\subsubsection{Self-reported self-regulation (7-point Likert scale)}
This self-report measure addresses RQ3. Participants reported their self-regulation in learning using three subscales from Pintrich et al.'s Motivated Strategies for Learning Questionnaire (MSLQ)~\cite{pintrich1991manual}. Specifically, we selected (1) three items measuring \textit{metacognitive self-regulation}, (2) three items measuring \textit{effort regulation}, and (3) one item measuring \textit{help seeking} (Table~\ref{appendix:self_reported_self_regulation_mslq}). Participants responded to each item on a 7-point Likert scale (1 = not at all true of me, 7 = very true of me), and we averaged item ratings within each subscale. Participants completed the questionnaire three times, in Stages 4, 7, and 10.

\subsubsection{Perceived support for the pathway to targeted help-seeking (7-point Likert scale)}
This self-reported measure helps answer RQ4 by assessing how helpful participants perceived \sysname{} to be for targeted help-seeking. In Stage 8, participants answered six questions about how well their learning environment (i.e., Figure~\ref{fig:interface}) supported the three metacognitive skills for targeted help-seeking. For example, one item asked whether the programming activities helped them recognize different knowledge scopes (e.g., v-for and v-bind) (1: not true at all, 7: very true). Figure~\ref{fig:result_metacognitive_skills} shows the full questionnaire.

\subsubsection{Learning experience (7-point Likert scale)}
To examine their learning experience with \sysname{}, \textit{HelpCoach} participants rated three Likert-scale items and provided open-ended explanations (Figure~\ref{fig:result_survey}) in Stage 8. The items assessed the system's usefulness for formulating specific questions, its disruption to problem-solving (RQ5), and participants' willingness to use it in the future.

\subsection{Results}

We begin by summarizing the main results of Study 2. Compared with \textit{Baseline}, \sysname{} significantly improved targeted help-seeking and led to greater knowledge retention one week later. High question specificity and participants' ratings of their intervention experience indicate that \sysname{} successfully supported the pathway to targeted help-seeking, leading to these outcomes. However, \sysname{} did not significantly outperform \textit{Baseline} in improving or retaining self-reported self-regulation, suggesting potential limits in transferring these skills to more general and independent settings, perhaps due to the short intervention period.

Before presenting the statistical results, we describe our analysis procedure. We used independent- and paired-samples t-tests for between- and within-condition comparisons of knowledge-test scores and question specificity, as Shapiro-Wilk tests showed their distributions did not significantly deviate from normality. For ordinal self-report measures, we used Mann–Whitney U and Wilcoxon signed-rank tests, respectively.

We confirmed that the \textit{HelpCoach} and \textit{Baseline} conditions did not differ significantly in pre-task knowledge (t-test, $t = 1.10, p = .280; M_{HelpCoach}=3.0, M_{Baseline}=3.5$), or pre-task self-reported \textit{metacognitive self-regulation} ($U = 197.5, p = .956; M_{HelpCoach}=4.8, M_{Baseline}=4.9$), \textit{effort regulation} ($U = 190.5, p = .806; M_{HelpCoach}=4.8, M_{Baseline}=4.9$), and \textit{help-seeking} ($U = 193.5, p = .859; M_{HelpCoach}=6.3, M_{Baseline}=6.1$). Throughout this section, we denote participants in the \textit{Baseline} and \textit{HelpCoach} conditions as B[1--20] and H[1--20], respectively, when reporting individual qualitative responses.

\subsubsection{\textbf{[RQ1] \sysname{} increased spontaneous question specificity during problem-solving.}}
\begin{figure}
    \centering
    \includegraphics[width=\linewidth]{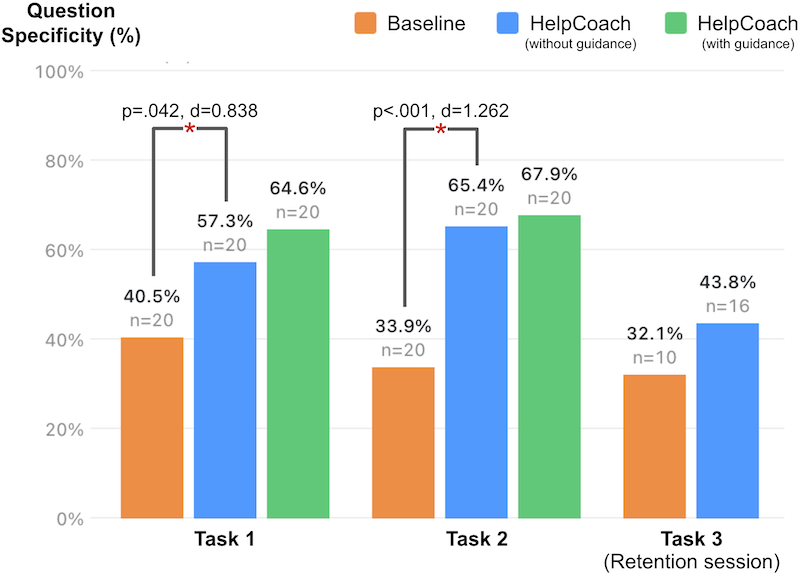}
    \caption{Question specificity by task and condition. Bars show mean question specificity, with sample sizes shown above each bar. \textit{HelpCoach} yielded significantly higher specificity in Task 1 and Task 2; no significant condition difference was observed in Task 3.}
    \Description{The grouped vertical bar chart displays three task groups on a scale from 0\% to 100\%, with horizontal grid lines every 20 percentage points. Orange bars represent Baseline, blue bars represent HelpCoach before revision, and green bars represent HelpCoach after revision. For Task 1, the means are 40.5\%, 57.3\%, and 64.6\%, respectively, each with n = 20. For Task 2, the corresponding means are 33.9\%, 65.4\%, and 67.9\%, each with n = 20. Thus, the green bar is tallest and the orange bar shortest in both groups. Task 3 contains only an orange Baseline bar at 32.1\% with n = 10 and a blue HelpCoach bar at 43.8\% with n = 16. Black brackets with red asterisks connect the orange and blue bars for Task 2, labeled p = .042, d = 0.838. Another black brackets with red asterisks connect the orange and blue bars for Task 2, labeled p < .001, d = 1.262. No brackets connect the blue and green bars, and no significance annotation appears for Task 3.}
    \label{fig:result_specificity}
\end{figure}

The results show that \sysname{} elicited the intended targeted help-seeking (Figure~\ref{fig:result_specificity}). To measure spontaneous question formulation, we analyzed participants' initial drafts before any \sysname{} intervention; each help-seeking attempt thus provided an opportunity for us to observe whether participants had learned to formulate a specific question independently. \textit{HelpCoach} participants asked a significantly higher proportion of specific questions than \textit{Baseline} participants in both Task 1 (t-test, $t = 2.11, p = .042; M_{HelpCoach} = 57.3\%, M_{Baseline} = 40.5\%$) and Task 2 ($t = 3.99, p < .001; M_{HelpCoach} = 65.4\%, M_{Baseline} = 33.9\%$). The effect was significant and large for both Task 1 and Task 2 ($d_{Task1} = 0.838, d_{Task2} = 1.262$)~\cite{lakens2013calculating}. Notably, more than half of the initial questions from \textit{HelpCoach} participants were specific in both tasks. Question specificity increased further after participants revised their questions with \sysname{}.

A week later, 30 of 40 participants returned to attend the retention sessions ($n_{HelpCoach}=17, n_{Baseline}=13$). Before analyzing question specificity in Task 3, we excluded conversations with fewer than three questions because they were too short to capture help-seeking behavior reliably; we excluded 4 of 30 conversations. \textit{HelpCoach} participants showed higher question specificity than \textit{Baseline} participants ($M_{\textit{HelpCoach}} = 43.7\%, M_{\textit{Baseline}} = 32.1\%$; Figure~\ref{fig:result_specificity}, Task 3), although the difference was not statistically significant (t-test, $t = 1.43, p = .166$). This pattern suggests a possible retention effect, but the smaller follow-up sample may have limited statistical power.

Participants reported that \sysname{} helped them ``build conceptual guidelines on [...] how to ask questions when [they] get stuck'' (H11) and become ``familiar with terms/phrases [...] to get the output [they] would like from a chatbot'' (H2, H6). H20 found the template useful for framing specific questions but questioned whether they could ``emulate this structure'' once the options were removed, noting that specifying knowledge components may require some prior understanding of the topic.

\subsubsection{\textbf{[RQ2] \textit{HelpCoach} improved knowledge retention.}}
\begin{figure}
    \centering
    \includegraphics[width=\linewidth]{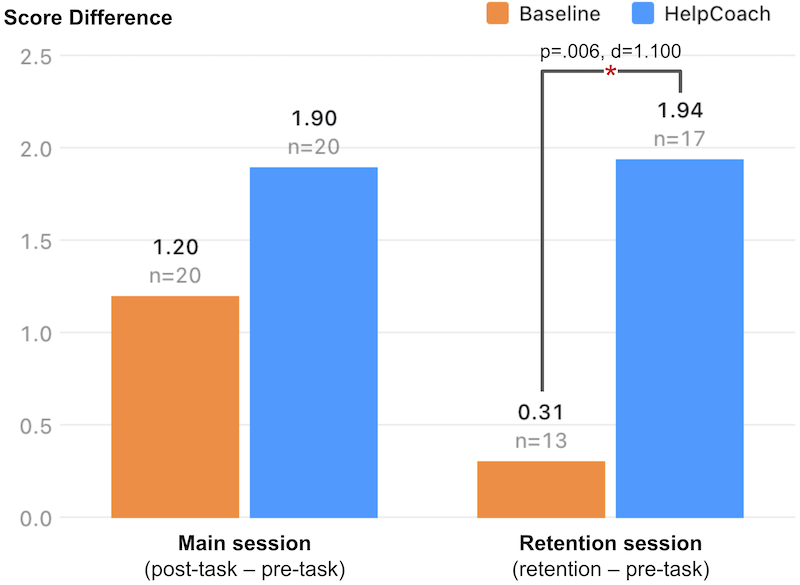}
    \caption{Knowledge gain by condition. Bars represent average knowledge-gain scores, with valid sample sizes shown above each bar. \textit{HelpCoach} produced significantly greater knowledge increase than \textit{Baseline} in the retention session.}
    \Description{The grouped vertical bar chart compares knowledge-gain scores across the main and retention sessions. HelpCoach is shown in blue and Baseline in orange. The vertical scale ranges from 0 to 2.5 in 0.5 increments. In the main session, the HelpCoach bar reaches a mean of 1.90 with 20 valid participants, whereas the Baseline bar reaches 1.20 with 20 participants. In the retention session, the HelpCoach mean is 1.94 with 17 participants, while the Baseline mean is 0.31 with 13 participants. The corresponding bracket is labeled p = .006 and d = 1.100 and includes a red asterisk. In both groups, the blue HelpCoach bar is visibly taller than the orange Baseline bar, with a larger separation between bars in the retention session.}
    \label{fig:result_knowledge_gain}
\end{figure}

In both conditions, participants showed significant knowledge gains from the pre- to post-knowledge tests (paired t-test, $t_{HelpCoach} = 4.87, p < .001$; $t_{Baseline} = 3.74, p = .001$), but the gains did not differ significantly between the conditions (t-test, $t = 1.39, p = .174; M_{\Delta HelpCoach}=1.9, M_{\Delta Baseline}=1.2$). However, knowledge retentions were significantly greater in \textit{HelpCoach} than in \textit{Baseline} ($t = 2.99, p = .006; M_{\Delta HelpCoach}=1.9, M_{\Delta Baseline}=0.3$; Figure~\ref{fig:result_knowledge_gain}). \textit{HelpCoach} participants maintained their knowledge gains, whereas \textit{Baseline} participants' gains declined. The effect of \sysname{} was large (Cohen's d, $d = 1.100$).

\subsubsection{\textbf{[RQ3] Self-reported self-regulation did not differ, but participants perceived self-regulatory benefits.}}

Pre-to-post changes in self-reported self-regulation did not differ significantly between conditions for any subscale: \textit{metacognitive self-regulation} ($U = 172.0, p = .442; M_{\Delta HelpCoach} = 0.3, M_{\Delta Baseline} = 0.1$), \textit{effort regulation} ($U = 187.5, p = .741; M_{\Delta HelpCoach} = 0.2, M_{\Delta Baseline} = 0.2$), or \textit{help-seeking} ($U = 154.0, p = .138; M_{\Delta HelpCoach} = -0.1, M_{\Delta Baseline} = 0.3$). Pre-to-retention changes also did not differ significantly for \textit{metacognitive self-regulation} ($U = 94.0, p = .496; M_{\Delta HelpCoach} = 0.5, M_{\Delta Baseline} = 0.4$), \textit{effort regulation} ($U = 104.0, p = .799; M_{\Delta HelpCoach} = 0.2, M_{\Delta Baseline} = 0.1$), or \textit{help-seeking} ($U = 102.5, p = .727; M_{\Delta HelpCoach} = -0.1, M_{\Delta Baseline} = -0.2$).

Despite the absence of significant differences in self-reports, \textit{HelpCoach} participants described perceived self-regulatory benefits. H10 noted that \sysname{} helped them ``keep the boundary of fighting our urges to still learn,'' despite the ease of requesting direct answers. H15 similarly reported that \sysname{} encouraged genuine learning with AI; otherwise, ``if it was ChatGPT, [they] would have just pasted the whole code and pasted the response.'' For H17, using \sysname{} prompted them to ``be careful to specify what [they] really want, so [they] can both achieve the goal of the assignment and learn things.''

Participants also perceived \sysname{} as helpful for developing self-regulation beyond the study sessions (Figure~\ref{fig:result_survey}, Q3) and expressed interest in using it for future learning (H1, H3). H10 described it as ``a great way of incorporating chat help in assignments in classrooms in the future'' and would use it to ``my benefit as opposed to an external chatbot when doing assignments to seek solutions in the middle ground, where I do not have to stay stuck, but I also do not have to forego learning to overcome it.'' H7, H9, and H13 similarly valued learning to use AI appropriately and avoiding ``being spoonfed answers in ways that were not conducive to learning.'' H18 noted that \sysname{} might be particularly useful for knowledge requiring ``reflection and critical thinking (e.g., algorithms).''

Participants valued discovering scaffold types they had not previously considered. H8 reported that \sysname{} ``gives suggestions that I have not thought of, which is helping me expand my ideas.'' H16 found both similar examples and step-by-step guidance useful for studying with AI and establishing foundational knowledge. These insights can inform their future learning more effectively.

\subsubsection{\textbf{[RQ4] \sysname{} supported the recognition and selection of knowledge components.}}

\begin{figure}
    \centering
    \includegraphics[width=\linewidth]{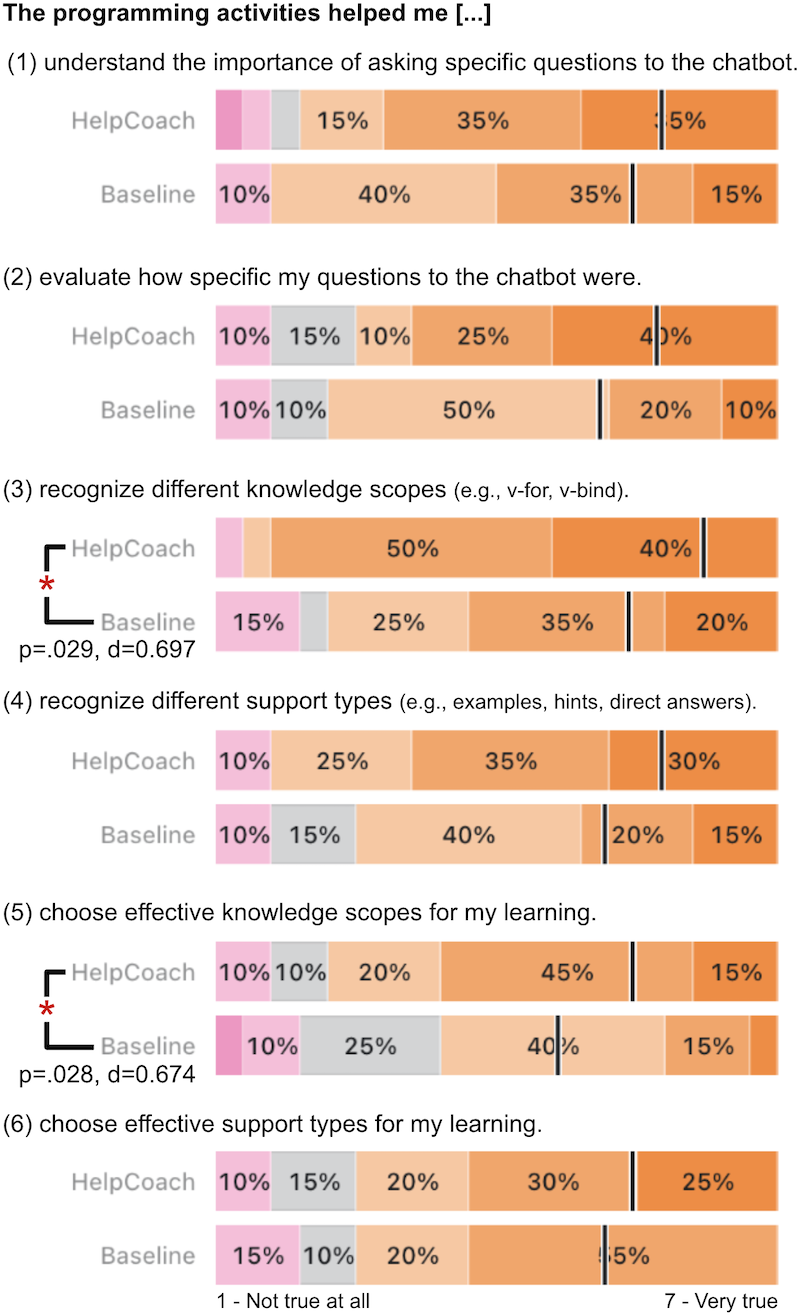}
    \caption{Perceived support for the pathway to targeted help-seeking by condition (20 participants each). The 100\% stacked bars show the distributions of responses on a seven-point scale (1 = not true at all; 7 = very true), and vertical black lines indicate mean ratings. Compared with \textit{Baseline}, \textit{HelpCoach} received significantly higher ratings for recognizing different knowledge components and choosing effective knowledge scopes.}
    \Description{The figure contains six vertically arranged pairs of 100\% stacked horizontal bars under the heading "The programming activities helped me […]." Each pair shows HelpCoach above Baseline. Responses progress from pink segments on the left, through gray and light orange, to darker orange on the right. The endpoints are labeled "1 – Not true at all" and "7 – Very true." Percentages are printed inside segments large enough to contain labels, and a vertical black line marks the mean of each bar. The six statements concern understanding the importance of specific questions, evaluating question specificity, recognizing different knowledge scopes, recognizing different support types, choosing effective knowledge scopes, and choosing effective support types. Across all six statements, the HelpCoach mean line appears to the right of the Baseline mean line, and its distributions generally contain larger dark-orange portions. Black comparison brackets with red asterisks appear for Statements 3 and 5. Their labels are p = .029, d = 0.697 and p = .028, d = 0.674, respectively. Statements 1, 2, 4, and 6 have no comparison brackets or significance annotations.}
    \label{fig:result_metacognitive_skills}
\end{figure}

Survey responses indicated that participants perceived \sysname{} as supporting knowledge-related metacognitive skills needed for targeted help-seeking (Figure~\ref{fig:result_metacognitive_skills}).

For the first skill, \numbercircled{1}~assessing question quality, \textit{HelpCoach} was rated higher than \textit{Baseline} for helping participants (1) understand the importance of asking specific questions (Mann-Whitney U test, $U = 155.0, p = .210; M_{HelpCoach} = 5.8, M_{Baseline} = 5.5$) and (2) evaluate the specificity of their own questions ($U = 139.0, p = .092; M_{HelpCoach} = 5.7, M_{Baseline} = 5.1$), but the differences were marginal.

Participants valued the continuous ratings of question specificity throughout the conversation (H1, H3, H8), ``motivating [them] to write better questions'' (H14). H19 described this feature as a safeguard that could ``pop up whenever [they] vibe code too much without really understanding what [they] are writing/copying and pasting,'' which they felt could prevent over-reliance on AI.

For the second skill, \numbercircled{2}~recognizing help options, \textit{HelpCoach} was significantly higher for helping participants (3) recognize different knowledge components ($U = 123.0, p = .029; M_{HelpCoach} = 6.2, M_{Baseline} = 5.4$) and marginally higher for helping them (4) recognize different support types ($U = 137.0, p = .080; M_{HelpCoach} = 5.7, M_{Baseline} = 5.2$). The corresponding effects were large ($d_3 = 0.697$) and moderate ($d_4 = 0.502$).

Participants valued seeing concrete options for formulating their questions (H6, H13). These options helped them ``realize which topic [they] are struggling on'' (H14), explore questions in an unfamiliar domain (H12, H13, H14), and ``restructure [their] question to better learn [...] instead of being fed the exact answer'' (H7).

For the third skill, \numbercircled{3}~choosing effective help-seeking strategies, \textit{HelpCoach} was rated significantly higher for helping participants (5) choose effective knowledge components ($U = 121.0, p = .028; d = 0.647$). However, ratings did not differ significantly between conditions for helping participants (6) choose effective scaffold types ($U = 170.0, p = .403$).

Participants reported that \sysname{} prompted them to consider ``what [they] wanted most from a set of assistance options'' (H2, H3) and to ``try taking small hints/examples, and even challenging [themselves] with questions'' (H19), rather than ``seeking an easy or inefficient answer'' (H10). It also helped them recognize that more specific questions can elicit better chatbot responses (H15). However, one participant requested more direct guidance for choosing among the three suggested options, suggesting a possible reason why \sysname{} did not significantly improve the selection of effective scaffold types.

\subsubsection{\textbf{[RQ5] \sysname{} prompted productive reflection with limited disruption.}}

\begin{figure}
    \centering
    \includegraphics[width=\linewidth]{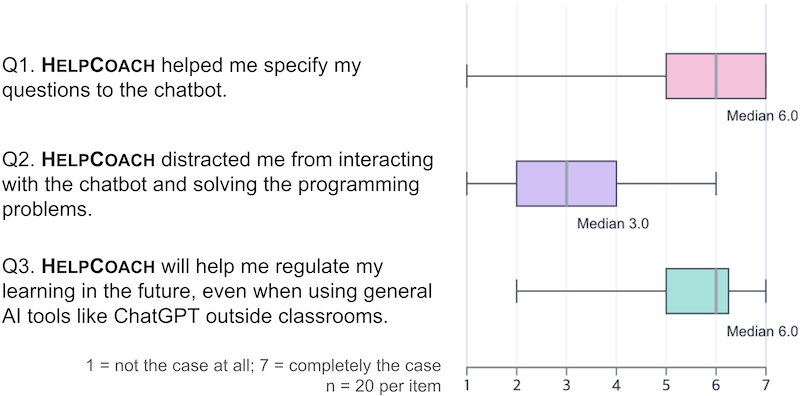}
    \caption{Distribution of \textit{HelpCoach} participants' responses to three survey items. \sysname{} was rated as helpful for specifying questions to the chatbot (median = 6) and for supporting future self-regulated learning with general AI tools (median = 6), whereas distraction from chatbot interaction and programming problem solving received a lower rating (median = 3).}
    \Description{Horizontal box plots summarize responses to three HelpCoach survey questions on a 1–7 scale. For Question 1, "HelpCoach helped me specify my questions to the chatbot," the median is 6; the middle half of ratings spans 5 to 7, and the full range spans 1 to 7. For Question 2, "HelpCoach distracted me from interacting with the chatbot and solving the programming problems," the median is 3; the middle half spans 2 to 4, and the full range spans 1 to 6. For Question 3, "HelpCoach will help me regulate my learning in the future, even when using general AI tools like ChatGPT outside classrooms," the median is 6; the middle half spans approximately 5 to 6.25, and the full range spans 2 to 7. Overall, participants reported stronger agreement with HelpCoach's question-specification and future self-regulation benefits than with its distracting effect.}
    \label{fig:result_survey}
\end{figure}

Participants generally did not find \sysname{} distracting (Figure~\ref{fig:result_survey}, Q2). They attributed this to the interaction taking ``only a few seconds to complete'' (H2), the options aligning with their intentions (H8), and the interface being simple to use (H6, H17). When \sysname{} did create cognitive friction, participants often found the pause productive: it clarified their intentions and directed them toward appropriate questions (H10, H11), helped them understand the problem scope (H4), and encouraged them to ask better questions (H14).

Some participants nevertheless experienced a brief learning curve (H1). H13 explained, ``At first, it was a little confusing how to use it. But it got a lot better very fast, and was intuitive to use, and was very helpful.'' Similarly, H15 initially did not recognize \sysname{} as a tool for refining questions, but found it useful once they understood its purpose.

When participants did feel distracted, the problem stemmed primarily from limited option coverage rather than intervention timing (H7, H12, H16, H20). H19 explained, ``Sometimes it was annoying to require a knowledge option and support type, especially when nothing really fit what I was asking for. Sometimes I just want a quick syntax question answered rather than a hint for a larger problem or an example.'' H5's response also suggests a potential expertise reversal effect~\cite{kalyuga2009expertise}: ``For those who already have knowledge in the relevant field, [\sysname{}] seems rather useless.'' These findings highlight the need for a scaffold that accommodates diverse help-seeking needs and quickly fades when students no longer need support.

%% file: section/70_discussion.tex
\section{Discussion}
We discuss possible explanations for the study results, directions for improving \sysname{}, and takeaways for designing human-AI interaction beyond learning contexts.

\subsection{Connecting Question Specificity and Knowledge Retention}

We hypothesized that asking specific questions would improve learning by eliciting only the help students need, thereby preserving opportunities for constructive engagement. Our results showed that \sysname{} increased question specificity and knowledge retention, suggesting more robust knowledge construction. We therefore examined whether more targeted chatbot responses could explain these benefits.

To assess whether the chatbot provided targeted help, we annotated the knowledge components and scaffold types in its responses using the chatbot response classifier developed for the help-response alignment evaluation (Section~\ref{sec:tech_eval_help_reponse_alignment}). We defined \textbf{targeted help} as a response that addressed a single knowledge component using a scaffold type effective for that component (Table~\ref{appendix:kc_type_classification}). For each participant, we calculated the proportion of chatbot responses that met these criteria.

Contrary to our expectations, the proportion of targeted help did not differ significantly between \textit{HelpCoach} and \textit{Baseline}. In Task 1, \textit{Baseline} participants received a higher proportion of targeted help on average ($M_{HelpCoach}=38.5\%, M_{Baseline}=42.7\%$), whereas \textit{HelpCoach} participants received a higher proportion in Task 2 ($M_{HelpCoach}=32.4\%, M_{Baseline}=24.5\%$). Neither difference was statistically significant (t-test, $t_{Task1}=0.584, p=.562; t_{Task2}=1.22, p=.229$). Thus, the increased question specificity in \textit{HelpCoach} did not translate into a higher proportion of targeted help.

This pattern was consistent with the relatively low response precision in both \textit{HelpCoach} ($M_{Knowledge}=.506, M_{Scaffold}=.367$) and \textit{Baseline} ($M_{Knowledge}=.546, M_{Scaffold}=.488$), with no significant differences between conditions. Several participants similarly reported receiving broader assistance than requested (B4, B7, B9, B20, H20). As one participant explained, ``the chatbot doesn't always match the kind of help I asked for, as there were times I asked for an example or hint, but it included the full answer in the reply. It is really helpful in aiding me to finish the assignment, but I feel like part of the learning process is skipped on my end.'' Precision was lower than in our technical evaluation ($precision_{Knowledge}=.770,\ precision_{Scaffold}=.607$). One possible explanation is that the longer conversational context in the study prompted the chatbot to incorporate information beyond the participant's immediate request.

\begin{figure*}
    \centering
    \includegraphics[width=0.75\linewidth]{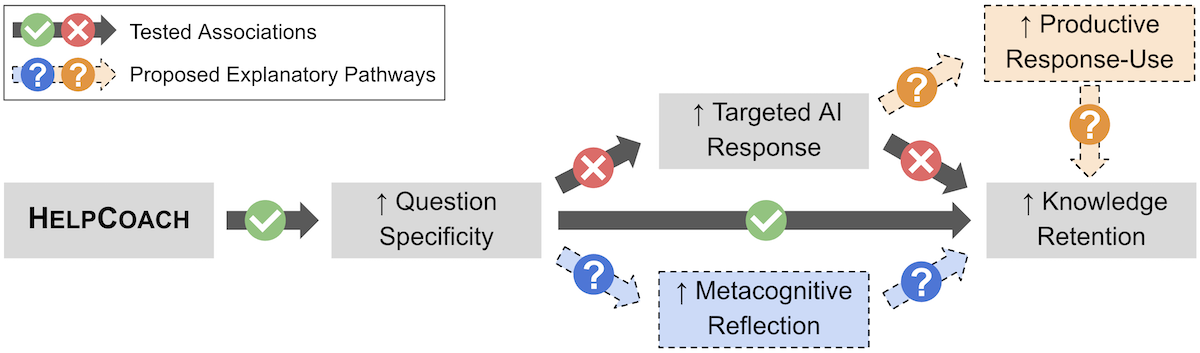}
    \caption{Tested associations and proposed pathways linking \sysname{} to knowledge retention. Solid gray arrows represent tested associations; green checkmarks indicate statistically significant associations, and red X marks indicate nonsignificant associations. \sysname{} was associated with greater question specificity, which was associated with greater knowledge retention. Dashed arrows with question marks show untested explanatory pathways through metacognitive reflection (blue) and response use (orange). These relationships should not necessarily be interpreted as causal effects.}
    \Description{A left-to-right conceptual diagram summarizes tested associations and two proposed explanations for how HelpCoach may relate to participants' knowledge retention. A legend in the upper-left corner explains the visual notation. Solid gray arrows represent tested associations. A green checkmark indicates a statistically significant association, while a red X indicates an association that was not statistically significant. Dashed colored arrows marked with question marks represent proposed explanatory pathways that have not been tested. Blue represents the metacognitive-reflection pathway, and orange represents the productive response-use pathway. In the main diagram, a box labeled "HelpCoach" appears on the far left. A solid gray arrow with a green checkmark connects HelpCoach to increased question specificity, indicating a statistically significant association. Increased question specificity is connected directly to increased knowledge retention by another solid arrow with a green checkmark. A diagonal solid arrow connects increased question specificity to an increased targeted AI response, but it has a red X, indicating the association was not statistically significant. A second solid arrow connects the targeted AI response to increased knowledge retention and also has a red X. Two untested explanations are shown with dashed arrows. The blue pathway proposes that question specificity may increase metacognitive reflection, which may subsequently improve knowledge retention. The orange pathway proposes that targeted AI responses may increase participants' use of chatbot responses, which may subsequently improve knowledge retention. Question marks on these pathways indicate proposed explanations rather than empirically tested findings.}
    \label{fig:discussion_association}
\end{figure*}

What, then, might explain the greater knowledge retention among \textit{HelpCoach} participants? We propose two alternative pathways. First, formulating specific questions may itself support learning (Figure~\ref{fig:discussion_association}, blue pathway). Doing so requires participants to reflect on what they do not understand, what assistance they need, and how that assistance should be delivered. This metacognitive process may trigger knowledge activation~\cite{brand2025prior, hattan2024leveraging} and help them connect knowledge components and construct more coherent knowledge structures. Second, participants may have engaged constructively with the chatbot's responses regardless of their scope (Figure~\ref{fig:discussion_association}, orange pathway). The proportion of targeted help was not significantly correlated with either knowledge gain (Spearman's $r=.076, p=.640$) or retention ($r=.288, p=.123$). Participants may therefore have learned by actively interpreting and applying even overly broad assistance, consistent with prior findings on constructive response use~\cite{jin2026reliancescope}.

Overall, \sysname{} increased both question specificity and knowledge retention, but our evidence does not establish that greater specificity directly caused the retention benefit. We also found no evidence that an increased proportion of targeted chatbot responses accounted for this effect. Future work should directly examine metacognitive reflection and response use as potential mediators of learning.

\subsection{Directions to Improve \sysname{}}

Participants' comments suggest several directions for alternative design choices for \sysname{}. First, the question-specificity classifier could better account for conversational context. H17 explained, ``I felt like [\sysname{}] could not use context clues. Like, I know saying `it' or `that' isn't specific, but it is nonetheless generally clear based on the preceding message.'' We intentionally classified pronoun-based questions as unspecified because prior literature treats such references as ambiguous~\cite{grasser1994}, and requiring students to clarify them may encourage metacognitive reflection. However, participants received \sysname{}'s revision prompts for 40\% of their questions, suggesting that the classifier may have been overly sensitive and caused unnecessary interruptions. We expect a trade-off: reducing its sensitivity may improve usability but weaken the metacognitive benefits of \sysname{}. Future designs could let instructors configure sensitivity, let students adjust it as part of their self-regulation, or adapt it automatically to students' performance as an additional fading mechanism.

Second, the revision template could be more flexible in accommodating diverse dialogue moves. As discussed in Section~\ref{sec:study1_qual_finding}, participants sometimes struggled to use the template for simple clarification questions (H9), final reviews of their entire code (H16), or follow-up questions about content introduced during the conversation, such as the chatbot's explanations or examples (H5). To avoid forcing the template into unsuitable contexts, we allowed participants to skip it when it did not fit their needs or when they were already confident about their question. \textit{HelpCoach} participants used this skip option for 8\% of their questions. Together with participants' generally favorable ratings, the low skip rate suggests that the template was not frequently disruptive. However, it may also indicate that the template discouraged follow-up or broader questions that could support learning in other ways. Future work could complement user-initiated skipping with system-initiated skipping or allow students to choose different template designs based on their question intents.

Lastly, researchers could explore different ways to represent and provide knowledge components in the template. Although we expanded our knowledge-component set while annotating 250 questions from Study 1, some participants felt that the options ``did not sufficiently cover all the different topics that [they] wanted to ask about'' (H20). Simply adding more predefined components may not resolve this limitation, as students may need to ask about relationships between components or select knowledge at finer or broader levels of granularity. We used a fixed number and granularity of knowledge components to simplify question revision and improve the accuracy of the scoped-exploration module. Alternative designs could represent knowledge components on a continuous canvas that students zoom in or out of, dynamically generate components at the desired granularity, or allow students to add their own components. Future work should investigate how these approaches can improve coverage and support flexible levels of knowledge specificity.

\subsection{Design Implications for Metacognition Scaffolding Systems}

\begin{figure}
    \centering
    \includegraphics[width=0.8\linewidth]{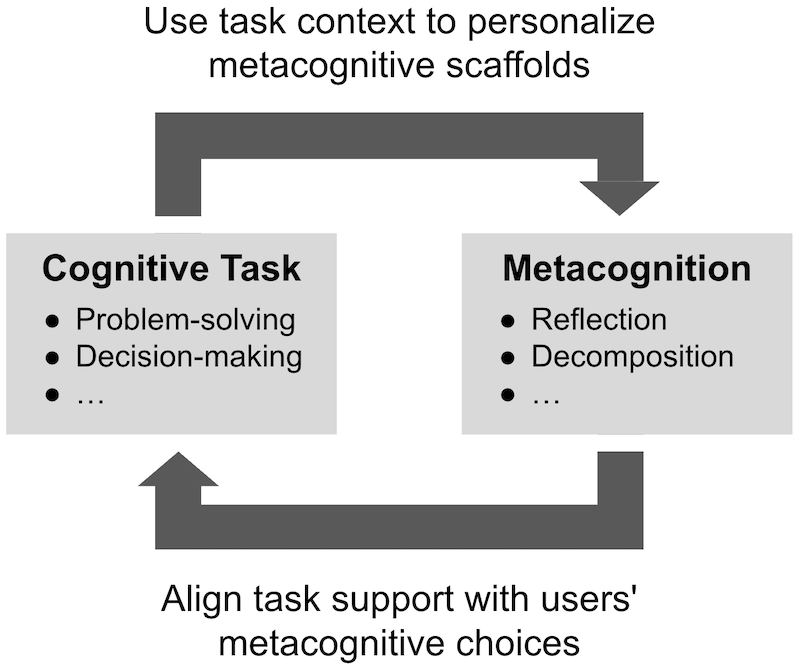}
    \caption{A loop for designing metacognitive scaffolds situated in cognitive tasks. A user's task context can inform personalized metacognitive scaffolds, while the user's metacognitive choices can determine the type and level of cognitive support a system provides. Together, these directions complete a loop that incentivizes users to reflect metacognitively during cognitive work.}
    \Description{The figure shows two light-gray boxes positioned side by side. The left box is titled "Cognitive Task" and lists "Problem-solving," "Decision-making," and an ellipsis indicating additional examples. The right box is titled "Metacognition" and lists "Reflection," "Decomposition," and an ellipsis. Above the boxes, the text reads, "Use problem-solving context to personalize metacognitive scaffolds." Beneath this text, a thick dark-gray bent arrow extends from the cognitive-task side toward the right and points downward to the metacognition box. Below the boxes, a second thick dark-gray bent arrow extends from the metacognition side toward the left and points upward to the cognitive-task box. The text beneath this arrow reads, "Align problem-solving support with users' metacognitive choices." The two arrows represent reciprocal information flow between cognitive-task support and metacognitive scaffolds.}
    \label{fig:discussion_design_implication}
\end{figure}

Supporting metacognition during cognitively demanding tasks is increasingly important in human-AI interaction across learning and knowledge work~\cite{xiao2025improving, tankelevitch2024metacognitive, willems2025use}. Metacognitive activities include monitoring one's understanding, calibrating confidence, and decomposing complex tasks into manageable subtasks~\cite{tankelevitch2024metacognitive}. These activities can support AI-assisted problem solving and decision making by helping users determine what to delegate and which strategies to adopt~\cite{jonassen2000toward}. A central design challenge is integrating metacognition into an ongoing task without disrupting the cognitive work it is intended to support. \sysname{} illustrates two principles for achieving this integration: (1) using task context to personalize metacognitive scaffolds and (2) aligning task support with users' metacognitive choices (Figure~\ref{fig:discussion_design_implication}). We present these principles as design implications for future metacognitive scaffolding systems.

\textit{Use task context to personalize metacognitive scaffolds.}
In situ scaffolds can draw on ongoing task context to infer users' immediate cognitive needs and reduce the effort required to shift from problem solving to reflection. \sysname{} assesses question specificity from students' requests and analyzes their code states and chat histories to recommend knowledge components relevant to their current work and difficulties. Recent research similarly uses contextual information, such as screenshots and interaction logs, to infer user needs and personalize support~\cite{lam2026just,jin2026thoughttrace,zhao2026behavior}. Future systems could develop richer cognitive user models that estimate learners' mastery of and engagement with individual knowledge components~\cite{lee2026knowsim,anderson1995cognitive}, adapting the timing, content, and granularity of metacognitive scaffolds accordingly.

\textit{Align task support with users' metacognitive choices.}
Metacognitive scaffolds should not end with reflection; users should see how their choices shape the support they subsequently receive. This feedback loop makes the benefits of deliberate reflection concrete and reinforces users' perceived agency over the task. \sysname{} closes this loop by passing the user-selected knowledge component and scaffold type to the chatbot, helping it provide the requested content and scaffold. Notably, users' needs and intentions may evolve throughout a task~\cite{kim2026discoverllm,zamfirescu2023johnny}; for example, in our work, students requested different knowledge components and scaffold types with each question. However, current AI systems struggle to follow such frequent, fine-grained changes in user requests~\cite{tack2026llms}. Future systems could employ stronger response constraints and explicit representations of users' current choices to keep AI support aligned with their evolving needs.

%% file: section/80_limitation.tex
\section{Limitations and Future Work}
We note several limitations of the current research. First, although \sysname{} and our findings may generalize to other problem-solving tasks, we avoid making strong generalizability claims because the system, its technical components, and our studies were designed specifically for web programming. Other domains may require different forms of in situ help-seeking scaffolds. Although our findings suggest that presenting knowledge-scaffold options can improve targeted help-seeking, we, as a community, need to understand how to operationalize the design for each new context. Future work could explore applications of \sysname{} to tasks such as open-ended writing and collaborative problem solving.

Second, we have limited evidence about how knowledge components should be represented for in situ scaffolds. Our internal pilot studies revealed a broad design space involving the granularity, abstraction, and relationships among components, but we evaluated only one predefined representation. Each dimension may affect the efficacy of \sysname{}: finer-grained components may support more targeted questions but also increase cognitive load. Moreover, no universally optimal representation may exist; effective representations may need to be generative and adapt dynamically to individual students. Future work should compare alternative representations and investigate whether AI-driven personalization can tailor them in real time.

Third, despite suggestive evidence of perceived benefits, our findings do not establish that \sysname{} significantly improves spontaneous targeted help-seeking and broader self-regulation in the longer term. The 50-minute exposure to \sysname{} may have been insufficient to produce measurable differences between conditions in these skills. Future longitudinal research, such as a semester-long classroom deployment, could determine whether the nonsignificant effects reflect limited exposure or a fundamental limitation of in situ scaffolds in fostering spontaneous self-regulation.

Fourth, although we examined help-seeking from question formulation through response reception, we did not investigate how students used the chatbot's responses for learning. Prior research suggests that constructive engagement with received help, such as understanding and applying AI-generated output, may compensate for poor help-seeking~\cite{aleven2016help,jin2026reliancescope}. Examining both the request and response-use sides of student-AI interaction may help explain why more specific questions did not translate into measurable long-term gains in self-regulation. Future work should investigate how help-seeking and response use jointly contribute to self-regulation development and whether in situ scaffolds can also promote constructive response-use behaviors.

%% file: section/90_conclusion.tex
\section{Conclusion}

GenAI is becoming a common source of assistance during problem-solving and is increasingly capable of providing correct, comprehensive answers. When AI can answer almost any question, knowing what help to ask for becomes a critical learning skill. This work introduced and evaluated \sysname{}, an in situ scaffold for such metacognitive self-regulation during programming problem solving. We found that \sysname{} could effectively prompt students to examine their knowledge gaps and articulate their needs through specific questions. Notably, our findings also provide preliminary evidence that scaffolding this help-seeking process may support more durable knowledge retention. This work highlights opportunities to use GenAI for contextual, real-time metacognitive scaffolding, and we envision GenAI that helps users not only solve the problems before them but also become more capable, reflective thinkers.

%% file: appendix/kc_type_classification.tex
% \usepackage{multirow}
\begin{table*}[ht]
\begin{tabular}{llll}
\Xhline{3\arrayrulewidth}
\textbf{Task} & \textbf{Knowledge Component}                    & \textbf{Type} & \textbf{Effective Scaffold Types}       \\
\Xhline{3\arrayrulewidth}
\multirow{7}{*}{Task 1} & {[}Data{]} Set up values in Vue data()               & Principles & {[}Hint{]}, {[}Example{]}, {[}Question{]} \\
              & {[}Loop{]} Show list items with v-for           & Facts         & {[}Steps{]}, {[}Explain{]}, {[}Check{]} \\
              & {[}Show data{]} Display Vue data with \{\{ \}\} & Facts         & {[}Steps{]}, {[}Explain{]}, {[}Check{]} \\
                        & {[}Objects/arrays{]} Build data with \{\} and {[}{]} & Rules      & {[}Check{]}, {[}Hint{]}, {[}Example{]}    \\
              & {[}HTML{]} Add bullet points and headings       & Rules         & {[}Check{]}, {[}Hint{]}, {[}Example{]}  \\
              & {[}Text{]} Slice and combine text               & Facts         & {[}Steps{]}, {[}Explain{]}, {[}Check{]} \\
              & {[}Errors{]} Understand console errors          & Rules         & {[}Check{]}, {[}Hint{]}, {[}Example{]}  \\ \hline
\multirow{8}{*}{Task 2} & {[}Input{]} Connect input to data with v-model       & Facts      & {[}Steps{]}, {[}Explain{]}, {[}Check{]}   \\
                        & {[}Methods{]} Write functions that change data       & Principles & {[}Hint{]}, {[}Example{]}, {[}Question{]} \\
              & {[}Click{]} Run a function on button click      & Facts         & {[}Steps{]}, {[}Explain{]}, {[}Check{]} \\
              & {[}Random{]} Make a random number               & Facts         & {[}Steps{]}, {[}Explain{]}, {[}Check{]} \\
              & {[}Text{]} Combine words and values             & Facts         & {[}Steps{]}, {[}Explain{]}, {[}Check{]} \\
              & {[}HTML{]} Add inputs, buttons, or attributes   & Rules         & {[}Check{]}, {[}Hint{]}, {[}Example{]}  \\
              & {[}Empty{]} Check if text is empty              & Facts         & {[}Steps{]}, {[}Explain{]}, {[}Check{]} \\
              & {[}Errors{]} Understand console errors          & Rules         & {[}Check{]}, {[}Hint{]}, {[}Example{]}  \\ \hline
\multirow{8}{*}{Task 3} & {[}CSS{]} Style cards with CSS properties            & Facts      & {[}Steps{]}, {[}Explain{]}, {[}Check{]}   \\
                        & {[}Grid{]} Arrange Bootstrap rows and columns        & Principles & {[}Hint{]}, {[}Example{]}, {[}Question{]} \\
              & {[}v-bind{]} Add a class conditionally          & Facts         & {[}Steps{]}, {[}Explain{]}, {[}Check{]} \\
              & {[}HTML{]} Structure tags and attributes        & Rules         & {[}Check{]}, {[}Hint{]}, {[}Example{]}  \\
              & {[}Loop{]} Show list items with v-for           & Facts         & {[}Steps{]}, {[}Explain{]}, {[}Check{]} \\
              & {[}Show data{]} Display Vue data with \{\{ \}\} & Facts         & {[}Steps{]}, {[}Explain{]}, {[}Check{]} \\
              & {[}Class{]} Apply a CSS class                   & Facts         & {[}Steps{]}, {[}Explain{]}, {[}Check{]} \\
              & {[}Errors{]} Understand console errors          & Rules         & {[}Check{]}, {[}Hint{]}, {[}Example{]}  \\ 
\Xhline{3\arrayrulewidth}
\end{tabular}
\caption{The list of knowledge components used for the studies and their corresponding effective scaffold types}
\label{appendix:kc_type_classification}
\end{table*}

%% file: appendix/scaffold_types.tex
\begin{table*}[ht]
\begin{tabular}{l}
\Xhline{3\arrayrulewidth}
\textbf{Scaffold Type}                           \\
\hline
{[}Steps{]} Tell me the answer or what to do     \\
{[}Explain{]} Explain or clarify a concept       \\
{[}Check{]} Review my work or idea, no fixes     \\
{[}Hint{]} Give clues, not the full answer       \\
{[}Example{]} Show a similar example, not answer \\
{[}Question{]} Ask questions to make me think   \\
\Xhline{3\arrayrulewidth}
\end{tabular}
\caption{The scaffold type used for the studies and for determining the specificity of a question.}
\label{appendix:scaffold_types}
\end{table*}

%% file: appendix/self_reported_self_regulation_mslq.tex
\begin{table*}[ht]
\begin{tabular}{ll}
\Xhline{3\arrayrulewidth}
\textbf{Subscale}     & \textbf{Item} \\ \hline
Metacognitive Self-Regulation &
  \begin{tabular}[c]{@{}l@{}}- I try to think through a topic and decide what I am supposed to learn \\ from it rather than just reading it over when studying.\\ - When studying in general, I try to determine which concepts I don't \\ understand well.\\ - When I study in general, I set goals for myself in order to direct my \\ activities in each study period.\end{tabular} \\ \hline
Effort Regulation &
  \begin{tabular}[c]{@{}l@{}}- I work hard to do well in general even if I don't like what I am doing.\\ - When the assignment is difficult, I give up or only study the easy \\ parts.(REVERSED)\\ - Even when course materials are dull and uninteresting, I manage to \\ keep working until I finish.\end{tabular} \\ \hline
Help-Seeking & \begin{tabular}[c]{@{}l@{}}- I ask an instructor or a chatbot to clarify concepts I don't understand \\ well.\end{tabular} \\
\Xhline{3\arrayrulewidth}
\end{tabular}
\caption{Participants answered the above questions on a 7-point Likert scale, where 1 = not at all true of me and 7 = very true of me.}
\label{appendix:self_reported_self_regulation_mslq}
\end{table*}

%% file: references.bib
@String{Computing = "Computing" }

@String{Computer = "{IEEE} Computer" }

@String{Academic = "Academic Press" }

@String{Springer = "Springer-Verlag" }

@article{reeve2011agency,
  title={Agency as a fourth aspect of students’ engagement during learning activities},
  author={Reeve, Johnmarshall and Tseng, Ching-Mei},
  journal={Contemporary educational psychology},
  volume={36},
  number={4},
  pages={257--267},
  year={2011},
  publisher={Elsevier}
}

@article{koedinger2012knowledge,
  title={The Knowledge-Learning-Instruction framework: Bridging the science-practice chasm to enhance robust student learning},
  author={Koedinger, Kenneth R and Corbett, Albert T and Perfetti, Charles},
  journal={Cognitive science},
  volume={36},
  number={5},
  pages={757--798},
  year={2012},
  publisher={Wiley Online Library}
}

@inproceedings{jin2026reliancescope,
author = {Jin, Hyoungwook and Yoo, Minju and Han, Jieun and Chen, Zixin and Ahn, So-Yeon and Wang, Xu},
title = {RelianceScope: An Analytical Framework for Examining Students' Reliance on Generative AI Chatbots in Problem Solving},
year = {2026},
isbn = {9798400722936},
publisher = {Association for Computing Machinery},
address = {New York, NY, USA},
url = {https://doi.org/10.1145/3774398.3811612},
doi = {10.1145/3774398.3811612},
booktitle = {Proceedings of the Thirteenth ACM Conference on Learning @ Scale},
pages = {136–147},
numpages = {12},
location = {Republic of Korea},
series = {L@S '26}
}

@inproceedings{hou2024effects,
author = {Hou, Irene and Mettille, Sophia and Man, Owen and Li, Zhuo and Zastudil, Cynthia and MacNeil, Stephen},
title = {The Effects of Generative AI on Computing Students’ Help-Seeking Preferences},
year = {2024},
isbn = {9798400716195},
publisher = {Association for Computing Machinery},
address = {New York, NY, USA},
url = {https://doi.org/10.1145/3636243.3636248},
doi = {10.1145/3636243.3636248},
booktitle = {Proceedings of the 26th Australasian Computing Education Conference},
pages = {39–48},
numpages = {10},
location = {Sydney, NSW, Australia},
series = {ACE '24}
}

@article{labadze2023role,
  title={Role of AI chatbots in education: systematic literature review},
  author={Labadze, Lasha and Grigolia, Maya and Machaidze, Lela},
  journal={International journal of Educational Technology in Higher education},
  volume={20},
  number={1},
  pages={56},
  year={2023},
  publisher={Springer}
}

@article{zimmerman2002becoming,
  title={Becoming a self-regulated learner: An overview},
  author={Zimmerman, Barry J},
  journal={Theory into practice},
  volume={41},
  number={2},
  pages={64--70},
  year={2002},
  publisher={Taylor \& Francis}
}

@article{bastani2025generative,
  title={Generative AI without guardrails can harm learning: Evidence from high school mathematics},
  author={Bastani, Hamsa and Bastani, Osbert and Sungu, Alp and Ge, Haosen and Kabakc{\i}, {\"O}zge and Mariman, Rei},
  journal={Proceedings of the National Academy of Sciences},
  volume={122},
  number={26},
  pages={e2422633122},
  year={2025},
  publisher={National Academy of Sciences}
}

@article{fan2025beware,
  title={Beware of metacognitive laziness: Effects of generative artificial intelligence on learning motivation, processes, and performance},
  author={Fan, Yizhou and Tang, Luzhen and Le, Huixiao and Shen, Kejie and Tan, Shufang and Zhao, Yueying and Shen, Yuan and Li, Xinyu and Ga{\v{s}}evi{\'c}, Dragan},
  journal={British Journal of Educational Technology},
  volume={56},
  number={2},
  pages={489--530},
  year={2025},
  publisher={Wiley Online Library}
}

@article{yan2024promises,
  title={Promises and challenges of generative artificial intelligence for human learning},
  author={Yan, Lixiang and Greiff, Samuel and Teuber, Ziwen and Ga{\v{s}}evi{\'c}, Dragan},
  journal={Nature human behaviour},
  volume={8},
  number={10},
  pages={1839--1850},
  year={2024},
  publisher={Nature Publishing Group UK London}
}

@article{nelson1981help,
  title={Help-seeking: An understudied problem-solving skill in children},
  author={Nelson-Le Gall, Sharon},
  journal={Developmental review},
  volume={1},
  number={3},
  pages={224--246},
  year={1981},
  publisher={Elsevier}
}

@article{chaiklin2003zone,
  title={The zone of proximal development in Vygotsky’s analysis of learning and instruction},
  author={Chaiklin, Seth and others},
  journal={Vygotsky’s educational theory in cultural context},
  volume={1},
  number={2},
  pages={39--64},
  year={2003},
  publisher={Cambridge, UK: Cambridge University Press.}
}

@article{schraw1998promoting,
  title={Promoting general metacognitive awareness},
  author={Schraw, Gregory},
  journal={Instructional science},
  volume={26},
  number={1},
  pages={113--125},
  year={1998},
  publisher={Springer}
}

@article{yoo2025teachers,
  title={How do teachers create pedagogical chatbots?: Current practices and challenges},
  author={Yoo, Minju and Jin, Hyoungwook and Kim, Juho},
  journal={arXiv preprint arXiv:2503.00967},
  year={2025}
}

@article{barcaui2025chatgpt,
  title={ChatGPT as a cognitive crutch: Evidence from a randomized controlled trial on knowledge retention},
  author={Barcaui, Andr{\'e}},
  journal={Social Sciences \& Humanities Open},
  volume={12},
  pages={102287},
  year={2025},
  publisher={Elsevier}
}

@article{pintrich1991manual,
  title={A manual for the use of the Motivated Strategies for Learning Questionnaire (MSLQ).},
  author={Pintrich, Paul R and others},
  year={1991},
  publisher={ERIC}
}

@article{newman1994adaptive,
  title={Adaptive help seeking: A strategy of self-regulated learning},
  author={Newman, Richard S},
  journal={Self-regulation of learning and performance: Issues and educational applications},
  pages={283--301},
  year={1994}
}

@incollection{pintrich2000role,
  title={The role of goal orientation in self-regulated learning},
  author={Pintrich, Paul R},
  booktitle={Handbook of self-regulation},
  pages={451--502},
  year={2000},
  publisher={Elsevier}
}

@article{nelson1986help,
  title={Help-seeking behavior in learning.},
  author={Nelson-Le Gall, Sharon},
  year={1986},
  publisher={ERIC}
}

@article{ryan2001avoiding,
  title={Avoiding seeking help in the classroom: Who and why?},
  author={Ryan, Allison M and Pintrich, Paul R and Midgley, Carol},
  journal={Educational Psychology Review},
  volume={13},
  number={2},
  pages={93--114},
  year={2001},
  publisher={Springer}
}

@article{miyake1979ask,
  title={To ask a question, one must know enough to know what is not known},
  author={Miyake, Naomi and Norman, Donald A},
  journal={Journal of verbal learning and verbal behavior},
  volume={18},
  number={3},
  pages={357--364},
  year={1979},
  publisher={Elsevier}
}

@inproceedings{hellas2023exploring,
author = {Hellas, Arto and Leinonen, Juho and Sarsa, Sami and Koutcheme, Charles and Kujanp{\"a}{\"a}, Lilja and Sorva, Juha},
title = {Exploring the Responses of Large Language Models to Beginner Programmers’ Help Requests},
year = {2023},
isbn = {9781450399760},
publisher = {Association for Computing Machinery},
address = {New York, NY, USA},
url = {https://doi.org/10.1145/3568813.3600139},
doi = {10.1145/3568813.3600139},
booktitle = {Proceedings of the 2023 ACM Conference on International Computing Education Research - Volume 1},
pages = {93–105},
numpages = {13},
location = {Chicago, IL, USA},
series = {ICER '23}
}

@inproceedings{kazemitabaar2024codeaid,
author = {Kazemitabaar, Majeed and Ye, Runlong and Wang, Xiaoning and Henley, Austin Zachary and Denny, Paul and Craig, Michelle and Grossman, Tovi},
title = {CodeAid: Evaluating a Classroom Deployment of an LLM-based Programming Assistant that Balances Student and Educator Needs},
year = {2024},
isbn = {9798400703300},
publisher = {Association for Computing Machinery},
address = {New York, NY, USA},
url = {https://doi.org/10.1145/3613904.3642773},
doi = {10.1145/3613904.3642773},
booktitle = {Proceedings of the 2024 CHI Conference on Human Factors in Computing Systems},
articleno = {650},
numpages = {20},
location = {Honolulu, HI, USA},
series = {CHI '24}
}

@InProceedings{gama2004metacognition,
author="Gama, Claudia",
editor="Lester, James C.
and Vicari, Rosa Maria
and Paragua{\c{c}}u, F{\'a}bio",
title="Metacognition in Interactive Learning Environments: The Reflection Assistant Model",
booktitle="Intelligent Tutoring Systems",
year="2004",
publisher="Springer Berlin Heidelberg",
address="Berlin, Heidelberg",
pages="668--677",
isbn="978-3-540-30139-4"
}

@article{bartholome2006matters,
  title={What matters in help-seeking? A study of help effectiveness and learner-related factors},
  author={Bartholom{\'e}, Tobias and Stahl, Elmar and Pieschl, Stephanie and Bromme, Rainer},
  journal={Computers in Human Behavior},
  volume={22},
  number={1},
  pages={113--129},
  year={2006},
  publisher={Elsevier}
}

@InProceedings{aleven2000limitations,
author="Aleven, Vincent
and Koedinger, Kenneth R.",
editor="Gauthier, Gilles
and Frasson, Claude
and VanLehn, Kurt",
title="Limitations of Student Control: Do Students Know when They Need Help?",
booktitle="Intelligent Tutoring Systems",
year="2000",
publisher="Springer Berlin Heidelberg",
address="Berlin, Heidelberg",
pages="292--303",
isbn="978-3-540-45108-2"
}

@article{wood1999help,
  title={Help seeking, learning and contingent tutoring},
  author={Wood, Heather and Wood, David},
  journal={Computers \& Education},
  volume={33},
  number={2-3},
  pages={153--169},
  year={1999},
  publisher={Elsevier}
}

@article{walter2017online,
  title={Online EEG-based workload adaptation of an arithmetic learning environment},
  author={Walter, Carina and Rosenstiel, Wolfgang and Bogdan, Martin and Gerjets, Peter and Sp{\"u}ler, Martin},
  journal={Frontiers in human neuroscience},
  volume={11},
  pages={286},
  year={2017},
  publisher={Frontiers Media SA}
}

@article{van2005cognitive,
  title={Cognitive load theory and complex learning: Recent developments and future directions},
  author={Van Merrienboer, Jeroen JG and Sweller, John},
  journal={Educational psychology review},
  volume={17},
  number={2},
  pages={147--177},
  year={2005},
  publisher={Springer}
}

@article{newman2002self,
  title={How self-regulated learners cope with academic difficulty: The role of adaptive help seeking},
  author={Newman, Richard S},
  journal={Theory into practice},
  volume={41},
  number={2},
  pages={132--138},
  year={2002},
  publisher={Taylor \& Francis}
}

@article{tuovinen1999comparison,
  title={A comparison of cognitive load associated with discovery learning and worked examples.},
  author={Tuovinen, Juhani E and Sweller, John},
  journal={Journal of educational psychology},
  volume={91},
  number={2},
  pages={334},
  year={1999},
  publisher={American Psychological Association}
}

@article{zhang2026does,
  title={How Does Learner Prior Language Knowledge Play in GenAI-Assisted Project-Based Learning for Creative Thinking?},
  author={Zhang, Zhihui and Chen, Cheng-Huan and Guo, Zihao and Chiu, Thomas KF},
  journal={Technology, Knowledge and Learning},
  pages={1--31},
  year={2026},
  publisher={Springer}
}

@article{raz2026knowledge,
  title={Knowledge reshapes inquiry by changing question asking ability and impacting academic assessment},
  author={Raz, Tuval and Kenett, Yoed N},
  journal={npj Science of Learning},
  volume={11},
  number={1},
  pages={19},
  year={2026},
  publisher={Nature Publishing Group UK London}
}

@InProceedings{aleven2004toward,
author="Aleven, Vincent
and McLaren, Bruce
and Roll, Ido
and Koedinger, Kenneth",
editor="Lester, James C.
and Vicari, Rosa Maria
and Paragua{\c{c}}u, F{\'a}bio",
title="Toward Tutoring Help Seeking",
booktitle="Intelligent Tutoring Systems",
year="2004",
publisher="Springer Berlin Heidelberg",
address="Berlin, Heidelberg",
pages="227--239",
isbn="978-3-540-30139-4"
}

@article{sawalha2024analyzing,
  title={Analyzing student prompts and their effect on ChatGPT’s performance},
  author={Sawalha, Ghadeer and Taj, Imran and Shoufan, Abdulhadi},
  journal={Cogent Education},
  volume={11},
  number={1},
  pages={2397200},
  year={2024},
  publisher={Taylor \& Francis}
}

@book{vygotsky1978mind,
  title={Mind in society: Development of higher psychological processes},
  author={Vygotsky, Lev Semenovich and Cole, Michael},
  year={1978},
  publisher={Harvard university press}
}

@article{pressiey1987cognitive,
  title={COGNITIVE STRA TEGIES},
  author={PressIey, Michael and Borkowski, John G and Schneider, Wolfgang},
  year={1987}
}

@article{king1991effects,
  title={Effects of training in strategic questioning on children's problem-solving performance.},
  author={King, Alison},
  journal={Journal of Educational psychology},
  volume={83},
  number={3},
  pages={307},
  year={1991},
  publisher={American Psychological Association}
}

@article{van2021connecting,
  title={Connecting teachers’ classroom instructions with children’s metacognition and learning in elementary school: van Loon MH et al.},
  author={van Loon, Mari{\"e}tte H and Bayard, Natalie S and Steiner, Martina and Roebers, Claudia M},
  journal={Metacognition and learning},
  volume={16},
  number={3},
  pages={623--650},
  year={2021},
  publisher={Springer}
}

@article{derry1986designing,
  title={Designing systems that train learning ability: From theory to practice},
  author={Derry, Sharon J and Murphy, Debra A},
  journal={Review of educational research},
  volume={56},
  number={1},
  pages={1--39},
  year={1986},
  publisher={Sage Publications Sage CA: Thousand Oaks, CA}
}

@Inbook{bull2010open,
author="Bull, Susan
and Kay, Judy",
editor="Nkambou, Roger
and Bourdeau, Jacqueline
and Mizoguchi, Riichiro",
title="Open Learner Models",
bookTitle="Advances in Intelligent Tutoring Systems",
year="2010",
publisher="Springer Berlin Heidelberg",
address="Berlin, Heidelberg",
pages="301--322",
isbn="978-3-642-14363-2",
doi="10.1007/978-3-642-14363-2_15",
url="https://doi.org/10.1007/978-3-642-14363-2_15"
}

@incollection{bull2013open,
  title={Open learner models as drivers for metacognitive processes},
  author={Bull, Susan and Kay, Judy},
  booktitle={International handbook of metacognition and learning technologies},
  pages={349--365},
  year={2013},
  publisher={Springer}
}

@inproceedings{brusilovsky2005engaging,
author = {Brusilovsky, Peter and Sosnovsky, Sergey},
title = {Engaging students to work with self-assessment questions: a study of two approaches},
year = {2005},
isbn = {1595930248},
publisher = {Association for Computing Machinery},
address = {New York, NY, USA},
url = {https://doi.org/10.1145/1067445.1067514},
doi = {10.1145/1067445.1067514},
booktitle = {Proceedings of the 10th Annual SIGCSE Conference on Innovation and Technology in Computer Science Education},
pages = {251–255},
numpages = {5},
location = {Caparica, Portugal},
series = {ITiCSE '05}
}

@article{bull2006computer,
  title={Computer-based formative assessment to promote reflection and learner autonomy},
  author={Bull, Susan and Quigley, Steven and Mabbott, Andrew},
  journal={engineering education},
  volume={1},
  number={1},
  pages={8--18},
  year={2006},
  publisher={Taylor \& Francis}
}

@article{mitrovic2007evaluating,
  title={Evaluating the effect of open student models on self-assessment},
  author={Mitrovic, Antonija and Martin, Brent},
  journal={International Journal of Artificial Intelligence in Education},
  volume={17},
  number={2},
  pages={121--144},
  year={2007},
  publisher={SAGE Publications Sage UK: London, England}
}

@InProceedings{shahrour2008does,
author="Shahrour, Gheida
and Bull, Susan",
editor="Nejdl, Wolfgang
and Kay, Judy
and Pu, Pearl
and Herder, Eelco",
title="Does 'Notice' Prompt Noticing? Raising Awareness in Language Learning with an Open Learner Model",
booktitle="Adaptive Hypermedia and Adaptive Web-Based Systems",
year="2008",
publisher="Springer Berlin Heidelberg",
address="Berlin, Heidelberg",
pages="173--182",
isbn="978-3-540-70987-9"
}

@article{sun2023effects,
  title={Effects of integrating an open learner model with AI-enabled visualization on students' self-regulation strategies usage and behavioral patterns in an online research ethics course},
  author={Sun, Jerry Chih-Yuan and Tsai, Hsueh-Er and Cheng, Wai Ki Rebecca},
  journal={Computers and Education: Artificial Intelligence},
  volume={4},
  pages={100120},
  year={2023},
  publisher={Elsevier}
}

@article{anderson1995cognitive,
  title={Cognitive tutors: Lessons learned},
  author={Anderson, John R and Corbett, Albert T and Koedinger, Kenneth R and Pelletier, Ray},
  journal={The journal of the learning sciences},
  volume={4},
  number={2},
  pages={167--207},
  year={1995},
  publisher={Taylor \& Francis}
}

@article{aleven2006toward,
  title={Toward meta-cognitive tutoring: A model of help seeking with a Cognitive Tutor},
  author={Aleven, Vincent and Mclaren, Bruce and Roll, Ido and Koedinger, Kenneth},
  journal={International journal of artificial intelligence in education},
  volume={16},
  number={2},
  pages={101--128},
  year={2006},
  publisher={SAGE Publications Sage UK: London, England}
}

@article{roll2007designing,
  title={Designing for metacognition—applying cognitive tutor principles to the tutoring of help seeking},
  author={Roll, Ido and Aleven, Vincent and McLaren, Bruce M and Koedinger, Kenneth R},
  journal={Metacognition and Learning},
  volume={2},
  number={2},
  pages={125--140},
  year={2007},
  publisher={Springer}
}

@article{roll2011improving,
  title={Improving students’ help-seeking skills using metacognitive feedback in an intelligent tutoring system},
  author={Roll, Ido and Aleven, Vincent and McLaren, Bruce M and Koedinger, Kenneth R},
  journal={Learning and instruction},
  volume={21},
  number={2},
  pages={267--280},
  year={2011},
  publisher={Elsevier}
}

@article{azevedo2005computer_a,
  title={Computer environments as metacognitive tools for enhancing learning},
  author={Azevedo, Roger},
  journal={Educational Psychologist},
  volume={40},
  number={4},
  pages={193--197},
  year={2005},
  publisher={Taylor \& Francis}
}

@incollection{azevedo2018using_b,
  title={Using hypermedia as a metacognitive tool for enhancing student learning? The role of self-regulated learning},
  author={Azevedo, Roger},
  booktitle={Computers as Metacognitive Tools for Enhancing Learning},
  pages={199--209},
  year={2018},
  publisher={Routledge}
}

@article{viberg2026efficiency,
  title={Efficiency vs. effectiveness: Self-regulated learning with LLM-mediated help-seeking},
  author={Viberg, Olga and Maggor, Yael Feldman and Wong, Jacqueline},
  journal={Learning Letters},
  volume={8},
  pages={60},
  year={2026},
  publisher={Public Knowledge Project}
}

@article{darvishi2024impact,
  title={Impact of AI assistance on student agency},
  author={Darvishi, Ali and Khosravi, Hassan and Sadiq, Shazia and Ga{\v{s}}evi{\'c}, Dragan and Siemens, George},
  journal={Computers \& Education},
  volume={210},
  pages={104967},
  year={2024},
  publisher={Elsevier}
}

@book{schunk1998self,
  title={Self-regulated learning: From teaching to self-reflective practice},
  author={Schunk, Dale H and Zimmerman, Barry J},
  year={1998},
  publisher={Guilford Press}
}

@inproceedings{kapoor2026exploring,
author = {Kapoor, Amanpreet and Denny, Paul and Porter, Leo and MacNeil, Stephen and Diaz, Marc},
title = {Exploring Student Behaviors and Motivations when using AI Teaching Assistants with Optional Guardrails},
year = {2026},
isbn = {9798400723520},
publisher = {Association for Computing Machinery},
address = {New York, NY, USA},
url = {https://doi.org/10.1145/3786228.3786233},
doi = {10.1145/3786228.3786233},
booktitle = {Proceedings of the 28th Australasian Computing Education Conference},
pages = {22–31},
numpages = {10},
location = {
},
series = {ACE '26}
}

@article{xiao2026transforming,
  title={Transforming GenAI Policy to Prompting Instruction: An RCT of Scalable Prompting Interventions in a CS1 Course},
  author={Xiao, Ruiwei and Ye, Runlong and Hou, Xinying and Wen, Jessica and Kumar, Harsh and Liut, Michael and Stamper, John},
  journal={arXiv preprint arXiv:2602.16033},
  year={2026}
}

@article{xiao2025improving,
  title={Improving student-AI interaction through pedagogical prompting: An example in computer science education},
  author={Xiao, Ruiwei and Hou, Xinying and Ye, Runlong and Kazemitabaar, Majeed and Diana, Nicholas and Liut, Michael and Stamper, John},
  journal={arXiv preprint arXiv:2506.19107},
  year={2025}
}

@article{ma2025should,
author = {Ma, Qianou and Peng, Weirui and Yang, Chenyang and Shen, Hua and Koedinger, Ken and Wu, Tongshuang},
title = {What Should We Engineer in Prompts? Training Humans in Requirement-Driven LLM Use},
year = {2025},
issue_date = {August 2025},
publisher = {Association for Computing Machinery},
address = {New York, NY, USA},
volume = {32},
number = {4},
issn = {1073-0516},
url = {https://doi.org/10.1145/3731756},
doi = {10.1145/3731756},
journal = {ACM Trans. Comput.-Hum. Interact.},
month = aug,
articleno = {41},
numpages = {27}
}

@book{jonassen1993structural,
  title={Structural knowledge: Techniques for representing, conveying, and acquiring structural knowledge},
  author={Jonassen, David H and Beissner, Katherine and Yacci, Michael},
  year={1993},
  publisher={Psychology Press}
}

@inproceedings{nielsen1994enhancing,
  title={Enhancing the explanatory power of usability heuristics},
  author={Nielsen, Jakob},
  booktitle={Proceedings of the SIGCHI conference on Human Factors in Computing Systems},
  pages={152--158},
  year={1994}
}

@article{van2010scaffolding,
  title={Scaffolding in teacher--student interaction: A decade of research},
  author={Van de Pol, Janneke and Volman, Monique and Beishuizen, Jos},
  journal={Educational psychology review},
  volume={22},
  number={3},
  pages={271--296},
  year={2010},
  publisher={Springer}
}

@article{chi2014icap,
  title={The ICAP framework: Linking cognitive engagement to active learning outcomes},
  author={Chi, Michelene TH and Wylie, Ruth},
  journal={Educational psychologist},
  volume={49},
  number={4},
  pages={219--243},
  year={2014},
  publisher={Taylor \& Francis}
}

@article{shute2008focus,
  title={Focus on formative feedback},
  author={Shute, Valerie J},
  journal={Review of educational research},
  volume={78},
  number={1},
  pages={153--189},
  year={2008},
  publisher={Sage Publications}
}

@incollection{clark2008cognitive,
  title={Cognitive task analysis},
  author={Clark, Richard E and Feldon, David F and Van Merrienboer, Jeroen JG and Yates, Kenneth A and Early, Sean},
  booktitle={Handbook of research on educational communications and technology},
  pages={577--593},
  year={2008},
  publisher={Routledge}
}

@article{schworm2012learning,
  title={e-Learning in universities: Supporting help-seeking processes by instructional prompts},
  author={Schworm, Silke and Gruber, Hans},
  journal={British Journal of Educational Technology},
  volume={43},
  number={2},
  pages={272--281},
  year={2012},
  publisher={Wiley Online Library}
}

@article{bai2026enhancing,
  title={Enhancing the effect of AI-assisted learning: the use of scaffolding strategies to develop students’ prompt engineering skills},
  author={Bai, Shurui and Yeung, Siu Sze and Lo, Chung Kwan},
  journal={Interactive Learning Environments},
  pages={1--22},
  year={2026},
  publisher={Taylor \& Francis}
}

@article{farrokhnia2024swot,
  title={A SWOT analysis of ChatGPT: Implications for educational practice and research},
  author={Farrokhnia, Mohammadreza and Banihashem, Seyyed Kazem and Noroozi, Omid and Wals, Arjen},
  journal={Innovations in education and teaching international},
  volume={61},
  number={3},
  pages={460--474},
  year={2024},
  publisher={Taylor \& Francis}
}

@article{chan2023students,
  title={Students’ voices on generative AI: Perceptions, benefits, and challenges in higher education},
  author={Chan, Cecilia Ka Yuk and Hu, Wenjie},
  journal={International journal of educational technology in higher education},
  volume={20},
  number={1},
  pages={43},
  year={2023},
  publisher={Springer}
}

@INPROCEEDINGS{feigenspan2012measuring,
  author={Feigenspan, Janet and Kästner, Christian and Liebig, Jörg and Apel, Sven and Hanenberg, Stefan},
  booktitle={2012 20th IEEE International Conference on Program Comprehension (ICPC)}, 
  title={Measuring programming experience}, 
  year={2012},
  volume={},
  number={},
  pages={73-82},
  doi={10.1109/ICPC.2012.6240511}
}

@inproceedings{ma2026not,
author = {Ma, Qianou and Koedinger, Kenneth R and Wu, Tongshuang},
title = {Not Everyone Wins with LLMs: Behavioral Patterns and Pedagogical Implications for AI Literacy in Programmatic Data Science},
year = {2026},
isbn = {9798400722783},
publisher = {Association for Computing Machinery},
address = {New York, NY, USA},
url = {https://doi.org/10.1145/3772318.3791283},
doi = {10.1145/3772318.3791283},
booktitle = {Proceedings of the 2026 CHI Conference on Human Factors in Computing Systems},
articleno = {139},
numpages = {22},
location = {
},
series = {CHI '26}
}

@article{graham2012measuring,
  title={Measuring and Promoting Inter-Rater Agreement of Teacher and Principal Performance Ratings.},
  author={Graham, Matthew and Milanowski, Anthony and Miller, Jackson},
  journal={Online Submission},
  year={2012},
  publisher={ERIC}
}

@article{yang2025analysing,
  title={Analysing nontraditional students' ChatGPT interaction, engagement, self-efficacy and performance: A mixed-methods approach},
  author={Yang, Mohan and Jiang, Shiyan and Li, Belle and Herman, Kristin and Luo, Tian and Moots, Shanan Chappell and Lovett, Nolan},
  journal={British Journal of Educational Technology},
  volume={56},
  number={5},
  pages={1973--2000},
  year={2025},
  publisher={Wiley Online Library}
}

@article{abbas2024harmful,
  title={Is it harmful or helpful? Examining the causes and consequences of generative AI usage among university students},
  author={Abbas, Muhammad and Jam, Farooq Ahmed and Khan, Tariq Iqbal},
  journal={International journal of educational technology in higher education},
  volume={21},
  number={1},
  pages={10},
  year={2024},
  publisher={Springer}
}

@article{stadler2024cognitive,
  title={Cognitive ease at a cost: LLMs reduce mental effort but compromise depth in student scientific inquiry},
  author={Stadler, Matthias and Bannert, Maria and Sailer, Michael},
  journal={Computers in Human Behavior},
  volume={160},
  pages={108386},
  year={2024},
  publisher={Elsevier}
}

@article{team2024learnlm,
  title={Learnlm: Improving gemini for learning},
  author={Team, LearnLM and Modi, Abhinit and Veerubhotla, Aditya Srikanth and Rysbek, Aliya and Huber, Andrea and Wiltshire, Brett and Veprek, Brian and Gillick, Daniel and Kasenberg, Daniel and Ahmed, Derek and others},
  journal={arXiv preprint arXiv:2412.16429},
  year={2024}
}

@incollection{kalyuga2009expertise,
  title={The expertise reversal effect},
  author={Kalyuga, Slava},
  booktitle={Managing cognitive load in adaptive multimedia learning},
  pages={58--80},
  year={2009},
  publisher={IGI Global Scientific Publishing}
}

@article{almeda2017help,
  title={Help avoidance: When students should seek help, and the consequences of failing to do so},
  author={Almeda, Victoria and Baker, Ryan and Corbett, Albert},
  journal={Teachers College Record},
  volume={119},
  number={3},
  pages={1--24},
  year={2017},
  publisher={SAGE Publications Sage CA: Los Angeles, CA}
}

@article{grasser1994,
  title={Question asking during tutoring},
  author={Graesser, Arthur C and Person, Natalie K},
  journal={American educational research journal},
  volume={31},
  number={1},
  pages={104--137},
  year={1994},
  publisher={Sage Publications}
}

@article{lee2026knowsim,
  title={KnowSim: Evaluating Information Calibration in LLM Assistants with User Simulators that Learn},
  author={Lee, Yoonjoo and Jin, Hyoungwook and Kim, Tae Soo and Zhang, Shaoyang and Laban, Philippe and Liao, Q. Vera},
  year={2026}
}

@article{bjork2013self,
  title={Self-regulated learning: Beliefs, techniques, and illusions},
  author={Bjork, Robert A and Dunlosky, John and Kornell, Nate},
  journal={Annual review of psychology},
  volume={64},
  number={1},
  pages={417--444},
  year={2013},
  publisher={Annual Reviews}
}

@article{avhustiuk2018illusion,
  title={The illusion of knowing in metacognitive monitoring: Effects of the type of information and of personal, cognitive, metacognitive, and individual psychological characteristics},
  author={Avhustiuk, Maria Mykolaivna and Pasichnyk, Ihor Demydovych and Kalamazh, Ruslana Volodymyrivna},
  journal={Europe's Journal of Psychology},
  volume={14},
  number={2},
  pages={317},
  year={2018}
}

@article{winne2002exploring,
  title={Exploring students’ calibration of self reports about study tactics and achievement},
  author={Winne, Philip H and Jamieson-Noel, Dianne},
  journal={Contemporary Educational Psychology},
  volume={27},
  number={4},
  pages={551--572},
  year={2002},
  publisher={Elsevier}
}

@article{azevedo2008externally,
  title={Why is externally-facilitated regulated learning more effective than self-regulated learning with hypermedia?},
  author={Azevedo, Roger and Moos, Daniel C and Greene, Jeffrey A and Winters, Fielding I and Cromley, Jennifer G},
  journal={Educational Technology Research and Development},
  volume={56},
  number={1},
  pages={45--72},
  year={2008},
  publisher={Springer}
}

@article{kruger1999unskilled,
  title={Unskilled and unaware of it: how difficulties in recognizing one's own incompetence lead to inflated self-assessments.},
  author={Kruger, Justin and Dunning, David},
  journal={Journal of personality and social psychology},
  volume={77},
  number={6},
  pages={1121},
  year={1999},
  publisher={American Psychological Association}
}

@article{chi1981categorization,
  title={Categorization and representation of physics problems by experts and novices},
  author={Chi, Michelene TH and Feltovich, Paul J and Glaser, Robert},
  journal={Cognitive science},
  volume={5},
  number={2},
  pages={121--152},
  year={1981},
  publisher={Elsevier}
}

@article{aleven2003help,
  title={Help seeking and help design in interactive learning environments},
  author={Aleven, Vincent and Stahl, Elmar and Schworm, Silke and Fischer, Frank and Wallace, Raven},
  journal={Review of educational research},
  volume={73},
  number={3},
  pages={277--320},
  year={2003},
  publisher={Sage Publications Sage CA: Thousand Oaks, CA}
}

@article{lakens2013calculating,
  title={Calculating and reporting effect sizes to facilitate cumulative science: a practical primer for t-tests and ANOVAs},
  author={Lakens, Dani{\"e}l},
  journal={Frontiers in psychology},
  volume={4},
  pages={863},
  year={2013},
  publisher={Frontiers Media SA}
}

@article{aleven2016help,
  title={Help helps, but only so much: Research on help seeking with intelligent tutoring systems},
  author={Aleven, Vincent and Roll, Ido and McLaren, Bruce M and Koedinger, Kenneth R},
  journal={International Journal of Artificial Intelligence in Education},
  volume={26},
  number={1},
  pages={205--223},
  year={2016},
  publisher={Springer}
}

@inproceedings{jin2025teachtune,
author = {Jin, Hyoungwook and Yoo, Minju and Park, Jeongeon and Lee, Yokyung and Wang, Xu and Kim, Juho},
title = {TeachTune: Reviewing Pedagogical Agents Against Diverse Student Profiles with Simulated Students},
year = {2025},
isbn = {9798400713941},
publisher = {Association for Computing Machinery},
address = {New York, NY, USA},
url = {https://doi.org/10.1145/3706598.3714054},
doi = {10.1145/3706598.3714054},
booktitle = {Proceedings of the 2025 CHI Conference on Human Factors in Computing Systems},
articleno = {1073},
numpages = {28},
location = {
},
series = {CHI '25}
}

@misc{openai2025college,
  title={Building anAI-Ready Workforce: A Look at College Student ChatGPT Adoption in the US},
  author={OpenAI},
  url={https://openai.com/global-affairs/college-students-and-chatgpt/},
  year={2025}
}

@misc{khan2025annual,
  title={2024-2025 Annual Report},
  author={Khan Academy},
  url={https://annualreport.khanacademy.org/},
  year={2025}
}

@inproceedings{lam2026just,
author = {Lam, Michelle S. and Shaikh, Omar and Xu, Hallie and Guo, Alice and Yang, Diyi and Heer, Jeffrey and Landay, James A. and Bernstein, Michael S.},
title = {Just-In-Time Objectives: A General Approach for Specialized AI Interactions},
year = {2026},
isbn = {9798400722783},
publisher = {Association for Computing Machinery},
address = {New York, NY, USA},
url = {https://doi.org/10.1145/3772318.3790713},
doi = {10.1145/3772318.3790713},
booktitle = {Proceedings of the 2026 CHI Conference on Human Factors in Computing Systems},
articleno = {802},
numpages = {26},
location = {
},
series = {CHI '26}
}

@article{jin2026thoughttrace,
  title={Thoughttrace: Understanding user thoughts in real-world llm interactions},
  author={Jin, Chuanyang and Li, Binze and Xie, Haopeng and Fang, Cathy Mengying and Li, Tianjian and Longpre, Shayne and Gu, Hongxiang and Chen, Maximillian and Shu, Tianmin},
  journal={arXiv preprint arXiv:2605.20087},
  year={2026}
}

@inproceedings{
kim2026discoverllm,
title={Discover{LLM}: From Executing Intents to Discovering Them},
author={Tae Soo Kim and Yoonjoo Lee and Jaesang Yu and John Joon Young Chung and Juho Kim},
booktitle={Forty-third International Conference on Machine Learning},
year={2026},
url={https://openreview.net/forum?id=uIi5FfjtwR}
}

@inproceedings{zamfirescu2023johnny,
author = {Zamfirescu-Pereira, J.D. and Wong, Richmond Y. and Hartmann, Bjoern and Yang, Qian},
title = {Why Johnny Can’t Prompt: How Non-AI Experts Try (and Fail) to Design LLM Prompts},
year = {2023},
isbn = {9781450394215},
publisher = {Association for Computing Machinery},
address = {New York, NY, USA},
url = {https://doi.org/10.1145/3544548.3581388},
doi = {10.1145/3544548.3581388},
booktitle = {Proceedings of the 2023 CHI Conference on Human Factors in Computing Systems},
articleno = {437},
numpages = {21},
location = {Hamburg, Germany},
series = {CHI '23}
}

@article{zhao2026behavior,
  title={Behavior Latticing: Inferring User Motivations from Unstructured Interactions},
  author={Zhao, Dora and Lam, Michelle S and Yang, Diyi and Bernstein, Michael S},
  journal={arXiv preprint arXiv:2604.07629},
  year={2026}
}

@article{tack2026llms,
  title={LLMs Get Lost in Evolving User Intent},
  author={Tack, Jihoon and Laban, Philippe and Neville, Jennifer},
  journal={arXiv preprint arXiv:2607.20734},
  year={2026}
}

@inproceedings{macina2023mathdial,
  title={Mathdial: A dialogue tutoring dataset with rich pedagogical properties grounded in math reasoning problems},
  author={Macina, Jakub and Daheim, Nico and Chowdhury, Sankalan and Sinha, Tanmay and Kapur, Manu and Gurevych, Iryna and Sachan, Mrinmaya},
  booktitle={Findings of the Association for Computational Linguistics: EMNLP 2023},
  pages={5602--5621},
  year={2023}
}

@article{tassoti2024assessment,
  title={Assessment of students use of generative artificial intelligence: Prompting strategies and prompt engineering in chemistry education},
  author={Tassoti, Sebastian},
  journal={Journal of Chemical Education},
  volume={101},
  number={6},
  pages={2475--2482},
  year={2024},
  publisher={ACS Publications}
}

@article{brand2025prior,
  title={Prior knowledge activation as preparation prior to instruction: does the coverage of relevant prior knowledge affect learning?},
  author={Brand, Charleen and Loibl, Katharina and Rummel, Nikol},
  journal={Instructional Science},
  volume={53},
  number={6},
  pages={1633--1661},
  year={2025},
  publisher={Springer}
}

@article{hattan2024leveraging,
  title={Leveraging what students know to make sense of texts: What the research says about prior knowledge activation},
  author={Hattan, Courtney and Alexander, Patricia A and Lupo, Sarah M},
  journal={Review of Educational Research},
  volume={94},
  number={1},
  pages={73--111},
  year={2024},
  publisher={Sage Publications Sage CA: Los Angeles, CA}
}

@inproceedings{tankelevitch2024metacognitive,
author = {Tankelevitch, Lev and Kewenig, Viktor and Simkute, Auste and Scott, Ava Elizabeth and Sarkar, Advait and Sellen, Abigail and Rintel, Sean},
title = {The Metacognitive Demands and Opportunities of Generative AI},
year = {2024},
isbn = {9798400703300},
publisher = {Association for Computing Machinery},
address = {New York, NY, USA},
url = {https://doi.org/10.1145/3613904.3642902},
doi = {10.1145/3613904.3642902},
booktitle = {Proceedings of the 2024 CHI Conference on Human Factors in Computing Systems},
articleno = {680},
numpages = {24},
location = {Honolulu, HI, USA},
series = {CHI '24}
}

@INPROCEEDINGS{willems2025use,
  author={Willems, Thijs and Khan, Sumbul and Huang, Qian and Camburn, Bradley and Sockalingam, Nachamma and Poon, King Wang},
  booktitle={2025 IEEE International Conference on Teaching, Assessment, and Learning for Engineering (TALE)}, 
  title={To Use or to Refuse? Re-Centering Student Agency with Generative AI in Engineering Design Education}, 
  year={2025},
  volume={},
  number={},
  pages={1-8},
  doi={10.1109/TALE66047.2025.11346653}
}

@article{jonassen2000toward,
  title={Toward a design theory of problem solving},
  author={Jonassen, David H},
  journal={Educational technology research and development},
  volume={48},
  number={4},
  pages={63--85},
  year={2000},
  publisher={Springer}
}

@article{jin2024codetree,
author = {Jin, Hyoungwook and Kim, Juho},
title = {CodeTree: A System for Learnersourcing Subgoal Hierarchies in Code Examples},
year = {2024},
issue_date = {April 2024},
publisher = {Association for Computing Machinery},
address = {New York, NY, USA},
volume = {8},
number = {CSCW1},
url = {https://doi.org/10.1145/3637308},
doi = {10.1145/3637308},
journal = {Proc. ACM Hum.-Comput. Interact.},
month = apr,
articleno = {31},
numpages = {37}
}
